\input harvmac.tex
\vskip 1.5in
\Title{\vbox{\baselineskip12pt 
\hbox to \hsize{\hfill}
\hbox to \hsize{\hfill CTP-SCU/2015013 }}}
{\vbox{
	\centerline{\hbox{Solutions in Bosonic String Field Theory
		}}\vskip 5pt
        \centerline{\hbox{and Higher Spin Algebras in $AdS$
		}} } }
\centerline{Dimitri Polyakov$^{}$\footnote{$^{\dagger(1),(2)}$}
{polyakov@sogang.ac.kr ; polyakov@scu.edu.cn; 
twistorstring@gmail.com
}}
\medskip
\centerline{\it Center for Theoretical Physics $^{(1)}$}
\centerline{\it College of Physical Science and Technology}
\centerline{\it Sichuan University, Chengdu 610064, China}
\centerline{\it }
\centerline{\it }
\centerline{\it Institute for Information Transmission Problems (IITP)$^{(2)}$}
\centerline{\it Bolshoi Karetny per. 19/1}
\centerline{\it 127994 Moscow, Russia}
\vskip .3in

\centerline {\bf Abstract}

We  find a class  of analytic solutions in  open bosonic string field theory,
parametrized by the chiral copy of higher spin algebra in $AdS_3$.
The solutions are expressed in terms 
of the generating function for the products of Bell 
polynomials
in derivatives of bosonic space-time coordinates $X^m(z)$ of the open string, 
which form is determined in this work. 
The products of these polynomials form a natural operator algebra realizations
of $w_\infty$ (area-preserving diffeomorphisms), 
enveloping algebra of SU(2) and higher spin algebra in
$AdS_3$. The class of SFT solutions found can, in turn, be interpreted as 
the ``enveloping of enveloping'', or the enveloping of $AdS_3$ higher spin algebra.
 We also discuss the extensions of this
class of solutions to superstring theory and their relations  to higher 
spin algebras
in higher space-time dimensions.

\Date{July 2015}

\vfill\eject

\lref\senf{A. Sen,  JHEP 9912 (1999) 027}
\lref\sens{A. Sen, B. Zwiebach, JHEP 0003 (2000) 002}
\lref\sent{L. Rastelli, A. Sen, B. Zwiebach, JHEP 0111 (2001) 035 }
\lref\vasilieveq{ M. A. Vasiliev, Phys. Lett. B 285 (1992) 225}
\lref\bek{ X. Bekaert, S. Cnockaert, C. Iazeolla, M. A. Vasiliev,
 hep-th/0503128 }
\lref\fvf{E.S. Fradkin, M.A. Vasiliev, Nucl. Phys. B 291, 141 (1987)}
\lref\fvs{E.S. Fradkin, M.A. Vasiliev, Phys. Lett. B 189 (1987) 89}
\lref\vcubic{M. A. Vasiliev, Nucl. Phys. B862 (2012) 341-408}
\lref\vmaf{M. A. Vasiliev, Sov. J. Nucl. Phys. 32 (1980) 439,
Yad. Fiz. 32 (1980) 855}
\lref\vmas{V. E. Lopatin and M. A. Vasiliev, Mod. Phys. Lett. A 3 (1988) 257}
\lref\vmulti{M. Vasiliev, Class.Quant.Grav. 30 (2013) 104006}
\lref\vmat{E.S. Fradkin and M.A. Vasiliev, Int. J. Mod. Phys. A 3 (1988) 2983}
\lref\fronsdal{C. Fronsdal, Phys. Rev. D18 (1978) 3624}
\lref\bbd{F. Berends, G. Burgers, H. Van Dam ,Nucl.Phys. B260 (1985) 295}
\lref\klebanov{ I. Klebanov, A. M. Polyakov,
Phys.Lett.B550 (2002) 213-219}
\lref\spinself{D. Polyakov, Phys.Rev.D82:066005,2010}
\lref\spinselff{D. Polyakov,Phys.Rev.D83:046005,2011}
\lref\selfsw{D. Polyakov, J. Phys. A46 (2013) 214012}
\lref\szw{A. Sen, B. Zwiebach, JHEP 0003 (2000) 002}
\lref\klebf{S. Giombi, I. Klebanov, arXiv:1308.2337}
\lref\erlf{T. Erler, JHEP 1311 (2013) 007}
\lref\erls{T. Erler, JHEP 1104 (2011) 107}
\lref\erlt{T. Erler, M. Schnabl, JHEP 0910 (2009) 066}
\lref\schnablf{M. Schnabl, Adv.Theor.Math.Phys. 10 (2006) 433-501}
\lref\schnabls{T. Erler, M.  Schnabl,JHEP 0910 (2009) 066}
\lref\schnablt{M. Kroyter, Y. Okawa, M. Schnabl, 
S. Torii, B. Zwiebach, JHEP 1203 (2012) 030}
\lref\kroyter{M. Kroyter, JHEP 1103 (2011) 081}
\lref\berkf{N. Berkovits, A. Sen, B. Zwiebach,  Nucl.Phys. B587 (2000) 147-178}
\lref\berks{N. Berkovits, Nucl. Phys. B450 (1995) 90}
\lref\berkt{N. Berkovits,  JHEP 0004 (2000) 022}
\lref\iaf{I. Arefeva, P. Medvedev, A. Zubarev, Mod. Phys. Lett. A6, 949
(1991)}
\lref\ias{I. Arefeva, A. Zubarev,  Mod.Phys.Lett. A8 (1993) 1469-1476}
\lref\iat{I. Arefeva, P. Medvedev, A. Zubarev, Nucl.Phys. B341 (1990) 464-498}
\lref\discf{I. Klebanov, A. Polyakov,  Mod.Phys.Lett. A6 (1991) 3273-3281}
\lref\discs{E. Witten, Nucl.Phys. B373 (1992) 187-213}
\lref\discself{D. Polyakov, Int.J.Mod.Phys. A22 (2007) 1375-1394}
\lref\rastelli{L. Rastelli, B. Zwiebach, JHEP 0109 (2001) 038}
\lref\witsft{E. Witten, Nucl.Phys. B268 (1986) 253}
\lref\witsfts{E. Witten, Phys.Rev. D46 (1992) 5467-5473}
\lref\yost{C. Preitschopf, C. Thorn and S. Yost, Nucl. Phys. B337 (1990) 363}
\lref\erler{T. Erler, JHEP 0801:013, (2008)}
\lref\soojong{M. Henneaux, S.-J. Rey,JHEP 1012 (2010) 007 }
\lref\barsf{I. Bars, Y. Matsuo, Phys.Rev. D66 (2002) 066003}
\lref\barss{I. Bars, I. Kishimoto,  Y. Matsuo, Phys.Rev. D67 (2003) 066002 }
\lref\barst{I. Bars, hep-th/0211238 }
\lref\klebanov{ I. Klebanov, A. M. Polyakov,
Phys.Lett.B550 (2002) 213-219}
\lref\giombif{S. Giombi, X. Yin, JHEP 1009 (2010) 115}
\lref\giombis{S. Giombi, X. Yin, Phys.Rev. D85 (2012) 086005}
\lref\maccf{L. Bonora, C. Maccaferri, D.D. Tolla, JHEP 1111 (2011) 107}
\lref\maccs{C. Maccaferri, JHEP 1405 (2014) 004}
\lref\macct{T. Erler, C. Maccaferri, JHEP 1410 (2014) 029}
\lref\rastf{L. Rastelli, B. Zwiebach, JHEP 0109 (2001) 038}
\lref\rasts{L. Bonora, C. Maccaferri, D. Mamone, 
M. Salizzone, hep-th/0304270}
\lref\doug{M. Douglas, H. Liu, G. Moore, B. Zwiebach, JHEP 0204 (2002) 022}
\lref\discf{I. Klebanov, A. Polyakov,  Mod.Phys.Lett. A6 (1991) 3273-3281}
\lref\discs{E. Witten, Nucl.Phys. B373 (1992) 187-213}
\lref\romans{C. Pope, L. Romans, X. Shen, Nucl. Phys. B339 (1990) 191}
\lref\selflast{D. Polyakov,  Phys.Rev. D90 (2014) 8, 086002}

\centerline{\bf  1. Introduction}

It is well-known that the equations of motion of Witten's cubic 
string field  theory
{\witsft, \witsfts}:
\eqn\grav{\eqalign{Q\Psi+\Psi\star\Psi=0}}
 resemble the Vasiliev's equations 
in the unfolding formalism in higher-spin theories 
~{\vasilieveq}
\eqn\lowen{dW+W\wedge\star{W}=0}
  (flatness condition for connection in
 infinite-dimensional higher-spin algebras)
 that determine the
interactions of the higher-spin gauge fields in this formalism, along with 
 equations for other master fields, containing higher-spin Weyl tensors  
and auxiliary fields  (see also e.g. ~{ \vmas, \fvs, \bek} 
for the  works/reviews on 
this remarkable formalism).
Higher spin holography strongly hints, 
however, that this resemblance may be much
more than just a formal similarity.
The generalized 1-form $W$ of (2) contains all the higher-spin gauge fields 
components in $AdS$ spaces which, by holography principle, are related
to various multi-index composite operators in the dual CFT's.
 Any of these CFT's, in turn, 
must be a low-energy limit of string theory in $AdS_{d+1}$, with the
$CFT_d$ correlators reproduced by the worldsheet correlation functions 
of the vertex operators in $AdS$ string theory, with the space-time
fields polarized along the boundary of the $AdS$ space.
On the other hand, the second-quantized string field $\Psi$, satisfying
the equation (1) is nothing but the expansion containing
 infinite number
of modes determined by these vertex operators.
Both string fields and higher spin gauge fields in the 
equations (1) and (2)  are known to be complicated objects to work with.
Despite the fact that the higher spin theories in $AdS$ spaces  can circumvent 
the restrictions imposed by the Coleman-Mandula's theorem, describing
the gauge-invariant higher-spin interactions is a highly nontrivial 
 problem since the gauge symmetry in these theories must be 
sufficiently powerful in order to eliminate unphysical degrees of freedom.
The restrictions imposed by such a gauge symmetry make the construction of the
interaction vertices in higher-spin theories a notoriously complicated problem.
While there was some progress in classification of the higher-spin 3-vertices
over recent years, the structure of the higher-order interactions 
(such as quartic interactions, presumably related to conformal 
blocks in  dual CFT's)
still remains obscure. The structure of these interactions is, 
however, crucial for our understanding of higher-spin extensions of the
holography principle and non-supersymmetric formulation of $AdS/CFT$.

At the same time, the string field theory still remains our best hope
 to advance towards background-independent formulation of string dynamics.
This, in turn, holds the keys to understanding 
string theories in curved backgrounds,
 such as AdS. Such 
 string theories are  also crucially
relevant to holography and gauge-string correspondence, however, 
little is known about them beyond the semiclassical limit .

Analytic solutions in string field theory appear to be one of the most crucial
ingredients  in order to
 approach such string theories in the SFT formalism, using the 
concept of background independence.
To illustrate this, suppose a string field $\Psi_0$ is a solution
of the equation (1). Then the form of (1) is invariant under the shift
\eqn\lowen{
\Psi\rightarrow{\tilde{\Psi}}=\Psi+\Psi_0}
with the simultaneous shift of the BRST charge
$Q\rightarrow{\tilde{Q}}$ , so that $Q^2={\tilde{Q}}^2=0$ and
the new nilpotent charge ${\tilde{Q}}$ defined according to
\eqn\lowen{{\tilde{Q}}\Psi=Q\Psi+\Psi_0\star\Psi+\Psi\star\Psi_0}
for any $\Psi$.
Then the new BRST charge ${\tilde{Q}}$ defines the new cohomology, different
from that of the original charge $Q$, corresponding to string theory in a
 new background, depending on the structure of $\Psi_0$. The advantage of 
this approach is that, in principle, it allows to
 explore the string theory in  new geometrical backgrounds
(e.g. in a curved geometry, such as AdS) while technically using
the operator products of the old string theory 
(say,  in originally flat background)
for the vertex operatos
in the new BRST cohomology, defined by ${\tilde{Q}}$.
This formalism is potentially more powerful than the first-quantized
formalism, which is background-dependent and where the vertex operator 
description is essentially limited to the flat space-time and 
semiclassical limit of curved backgrounds.
Unfortunately, however, the major obstacle is that 
identifying  analytic solutions
of the equation (1) is 
hard because of the complexity of the star product
in (1). For this reason, there are not  many known examples of 
analytic solutions having
a clear physical interpretation.
 One of the most fascinating 
and well-known solutions, describing
the nonperturbative tachyonic vacuum in string theory is of
course the class of the Schnabl's solutions {\schnablf}, later generalized
in a number of important papers, in particular, such as
{\schnabls, \schnablt, \erlf,\erls, \erlt} 
 which were discovered several years
ago and in particular
 used to prove the Sen's conjecture ~{\senf, \sens} Since that remarkable
paper by Schnable ~{\schnablf} there were 
many other interesting works describing the related
SFT  solutions, both in cubic theory and in Berkovits
SFT theory {\erlf, \erls, \erlt, \berkf, \maccf, \maccs, \macct}
such as algebraic SFT solutions, the analytic solutions describing various 
nonperturbative
processes such as D-brane translations. Despite that,
 classes of the SFT solutions , relevant to particular geometric 
backgrounds in string theory ,
 in particular those that would allow us to advance towards
consistent formulation of string theories in different space-time geometries,
 still mostly remain beyond our reach. 
One reason for this is that the star product in the 
equation
(1) is hard to  work with in practice
~{\witsft, \witsfts, \yost, \iat, \ias, \vmat, \berks,
\berkf, \berkt, \rastelli, \rastf, \rasts, \doug }.
 In general, this product is quite different
 from the conventional  Moyal product or the product in the Vasiliev's
equations (2), although for certain restricted classes of string fields
the star product of (1) can be mapped to the Moyal product
{\barsf, \barss, \barst}.
In general, however, the star product involves the global conformal '
transformations
\eqn\grav{\eqalign{
f_k^N(z)=e^{{2i\pi(k-1)}\over{N}}({{1-iz}\over{1+iz}})^{2\over{N}}}}

 that map the string fields living on separate worldsheets
to $N$ wedges of a single disc. 
The behavior of generic string fields (containing all sorts of 
off-shell non-primary operators) under such global conformal transformations
 easily wobbles out of control beyond 
any low-level truncation, making it hard to evaluate
the star product by straightforward computation of the correlators in OSFT.
There  are very few known  exceptions to that, 
such as the wedge states  or the
 special degenerate case of 
$\Psi$ constrained to primaries and their derivatives However, such fields 
form  too small a subset in the space of all the operators. 
The known SFT solutions
constrained to this subset do exist. However, 
with the exception of the Schnabl-related 
class of solutions,
they are typically irrelevant to non-perturbative background deformations
(see the discussion in the next section). 
At the same time, there exists a sufficiently large  class of the operators 
(far larger than the class of the primary fields) which behaves in
a rather compact and controllable way under (5), forming a closed 
subset of operators under the global conformal transformations.
Typically, these operators have the form
\eqn\grav{\eqalign{T^{(N)}=\sum_{k=1}^N\sum_{N|N_1...N_k}
\lambda^{(N)}_{N_1...N_k}B^{(N_1)}...B^{(N_k)}}}
with the sum taken over the partitions of total
conformal dimension $N$ of  $T^{(N)}$
and with $B^{(N_i)}(\partial{X},\partial^2{X},...,\partial^{N_j}{X})$ 
being the Bell polynomials of rank $N_j$ in the worldsheet derivatives
of string or superstring space-time coordinates or the ghost fields
(we shall review the basic properties of these objects in the next section).
The structure of the correlators of the operators of the form (6)
, analyzed in
 this work, as well as their transformation properties under (5)
makes them natural candidates to test for the analytic solutions of (1).
At the same time, it turns out that the
structure  constants of higher spin algebras in $AdS$ can be
realized in terms of the OPE structure constants 
of the operators of the type (6). This makes a natural guess that
the SFT solutions of the form (6) describe backgrounds 
with nonperturbative higher-spin configurations stemming from 
 full interacting (to all orders) higher-spin theory in $AdS$. 
More precisely, this means the 
following. Suppose that somehow we manage to take a glimpse
into full consistently interacting higher-spin theory and 
the higher spin interactions to all orders. Of course the Lagrangian 
of such a theory would be immensely complex, with all 
due restrictions imposed by the gauge invariance, 
with nonlocalities etc. One would also 
expect issues with unitarity  as well, at least in backgrounds other than AdS. 
Assume, however, that we managed to identify such a higher-spin action and 
to solve the equations of motion, i.e. to find the higher-spin configuration 
minimizing this action. From the string theory point of view, such a
 background would correspond to a certain conformal fixed point, with vanishing
$\beta$-functions of higher-spin vertex operators. An attempt
 to compute such $\beta$-functions straightforwardly would be hopeless,
 since that would require
summing up contributions from all orders of the string perturbation theory.
However, instead of computing the  $\beta$-function, one can try to find
an analytic solution describing the shift
$Q\rightarrow{\tilde{Q}}_{HS}$ from the flat background to the one involving the
nonperturbative higher spin configuration in AdS. To make a parallel to the
Schnabl's solution for
nonperturbative tachyonic background  note that,
from the  on-shell string theory point of view, this solution 
describes the minimum 
of the tachyon potential stemming the tachyon's $\beta$-function,
 computed to all orders of the string perturbation theory.
Given the SFT solution for nonperturbative higher-spin
background, the cohomology of ${\tilde{Q}}_{HS}$ would then describe the
physical properties of such a background.
At the first glance, the structure of such a solution 
must be enormously complicated. Nevertheless, let's try to imagine its possible
structure. The complete fully interacting higher-spin theory in $AdS_d$,
 no matter
how complicated its Lanrangian might be, is largely determined by 
two objects: structure constants of the higher-spin algebras in $AdS$
and conformal blocks in the dual $CFT_{d-1}$. 
Moreover, as we shall argue in the next section, as
 far as the cubic SFT is concerned, for substantially large class of solutions
the structure constants of the higher-spin algebra
 (more precisely, the enveloping
of this algebra) 
alone constitute a sufficient information to
 control the solutions we are looking for.
 Thus, if one is
able to find a class of
SFT solutions determined by the structure constants of the HS algebra,
this already would be a strong signal that it describes the higher spin background
we are interested in. The rest of this paper is organized as follows.
In the section 2 we shall discuss, as a warm-up example, a set of simple 
SFT solutions that involve the primary fields only and describe the
$perturbative$ background deformations. Remarkably, one particular example of 
these solutions is given by the discrete states in $c=1$ model where both the
structure constants of the $AdS_3$ higher-spin algebra appear and the vertex 
operators are described in terms of products of the 
Bell polynomials of the type (4). In the section 3 we develop
 the OPE formalism for the Bell polynomials of string fields, 
evaluating their structure constants.  We find that these structure constants
can be obtained from simple generating function $G(x,y)$ of two variables, which
 series expansion is determined by coefficients
related to $AdS_3$ structure constants. Next, we propose an ansatz
of the form (6) solving (1). The solution is given by the certain composite 
function $F(G)$ satisfying certain defining relations 
derived in this paper and structurally
can be thought of as an envelopping of the higher-spin algebra.
In the concluding section we discuss the physical implications of our
results and the generalizations relating analytic OSFT solutions to
higher-spin algebras in higher dimensional $AdS$ spaces.

\centerline{\bf 2. Structure Constants, Higher Spins and SFT solutions:
 a warm-up example}

One particularly simple and almost obvious example of a class of string fields
solving (1) can be constructed as follows.

Let $V_{i}(z,p) (i=1,...)$ be the set of all physical vertex operators
in string theory in the cohomology of the original BRST charge $Q$
(primary fields of ghost number 1 and conformal dimension 0)
and $\lambda^i(p)$ are the corresponding space-time fields
(where $p$ is the momentum in space-time and we suppress the space-time 
indices for the brevity).
Then the string field
\eqn\grav{\eqalign{\Psi_0=\sum_i\lambda^iV_i}}
is the solution of (1) provided that the zero $\beta$-function
conditions:
\eqn\grav{\eqalign{\beta_{\lambda^i}=0}}  are imposed on the space-time fields 
in the leading order of the perturbation theory.
This statement is easy to check.
Indeed, the on-shell invariance conditions on $V_i$ imply
$\lbrace{Q},\lambda^i{V_i}\rbrace={\hat{L}}\lambda^i=0$
where ${\hat{L}}$ is some differential operator (e.g. a Laplacian
plus the square of mass) acting on $\lambda^i$.
Next, since the operators are the dimension zero primaries, they are invariant
 under the transformations (5) and therefore the star product can be computed
simply by using
\eqn\grav{\eqalign{<<\Psi,\Psi\star\Psi>>=
 <\prod_{n=1}^3f_n^3\circ\Psi(0)>=\sum_{i,j,k}C_{ijk}\lambda^i\lambda^j\lambda^k}}
where $C_{ijk}(p_1,p_2)$ are the structure constants  in front of the 
simple pole in the OPE of the vertex operators:
\eqn\lowen{V_i(z_1,p_1)V_j(z_2,p_2)\sim(z_1-z_2)^{-1}{C_{ijk}(p_1,p_2)}
V^k({1\over2}(z_1+z_2),p_1+p_2)}
Substituting  (7), (9) into SFT equations of motion (1)
then leads to the constraints on $\lambda^i$ space-time fields:
\eqn\grav{\eqalign{
{\hat{L}}\lambda^i+C^i_{jk}\lambda^j\lambda^k=0}}
which are nothing but $\beta_{\lambda^i}=0$ equation (8) in the leading order.
Note that the SFT solution (7) is entirely fixed by the leading order
contribution to the $\beta$-function (which are completely determined
by the 3-point correlation functions of the vertex operators)
and does not depend on the higher-order corrections (related to the higher-point
correlators). The higher order corrections to the $\beta$-function only appear
upon the deformation (4) of the BRST charge related to the 
solution (7) which, in this case, simply reduces to
$Q\rightarrow{\tilde{Q}}=Q+\sum{\lambda_i{V^i}}$.
The 4-point functions of the $V_i$ vertex operators will then determine the
solution of the equation (1) with $Q$ replaced by ${\tilde{Q}}$. This, in turn,
will lead to the further shift of ${\tilde{Q}}$ in the next order etc.,
so the whole procedure can be performed order by order.
The physical meaning of these deformations is also quite clear:
they define, order by order, the $perturbative$ changes of the background
caused by the RG flows from the original conformal point (corresponding
to flat background) to the new fixed point (corresponding to a certain 
solution of the low-energy effective equations of motion).
Physically, far more interesting is of course the case 
when the operators entering (7) are no longer the primaries of any fixed
conformal dimensions and are off-shell, but still solve (1) with the constraints
of the type (11).
Then, $\lambda$ describes the
 background which is beyond the reach
of the conventional string perturbation theory while the $C$-constants describe
the new $2d$ CFT related to this $non-perturbative$ background change.
This is precisely the type of the higher-spin related SFT  solution 
we will be looking for. The instructive point here (which follows from 
the above discussion) 
is that, if we start with the SFT equation (1) with the
unperturbed BRST charge of the bosonic theory:
\eqn\grav{\eqalign{Q=\oint{dz}\lbrace{cT-bc\partial{c}}\rbrace=
\oint{dz}{\lbrace}-{1\over2}\partial{X_m}\partial{X^m}+bc\partial{c}\rbrace}}
the higher-spin solution we are searching for shouldn't depend on higher-point
correlators or the conformal blocks, but only on the structure
constants of the higher-spin algebra.
The final remark we shall make before moving further regards the appearance
of the higher spin algebra in the SFT solution of the type (7) at the
$perturbative$ level, as well as the appearance of the Bell polynomials
as the operators realizing this algebra.
Consider the
noncritical open one-dimensional bosonic string theory (also known as the $c=1$
model). It is well-known that this string theory does not contain a photon
in the massless spectrum, however, due to the $SU(2)$ symmetry at the self-dual
point, it does contain the $SU(2)$ multiplet of the discrete states which
are physical at integer or half-integer momentum values only and become massless
upon the Liouville dressing. To obtain the vertex operators for these states,
consider the $SU(2)$ algebra generated by
\eqn\grav{\eqalign{
T_{\pm}=\oint{dz}e^{\pm{i}X{\sqrt{2}}}\cr
T_0={i\over{\sqrt{2}}}\partial{X}}}
where $X$ is a single target space coordinate and the dressed 
BRST-invariant highest
weight vector
\eqn\grav{\eqalign{V_l=\int{dz}e^{(ilX+(l-1)\varphi){\sqrt{2}}}}}
where $\varphi$ is the Liouville field and $l$ is integer or half-integer.
The SU(2) multiplet of the operators is then obtained by repeatedly
acting on $V_l$
with the lowering operator $T_{-}$ of SU(2):
\eqn\grav{\eqalign{U_{l|m}=T_{-}^{l-m}V_l\cr
-l\leq{m}\leq{l}
}}
The dressed $U_{l|m}$ operators are the physical operators (massless states)
of the $c=1$ model and are the worldsheet integrals of primary fields of 
dimension one (equivalently, the primaries of dimension $0$ at the unintegrated 
$b-c$ ghost number 1 picture)

Manifest expressions for the {$U_{l|m}$} vertex operators are complicated,
however, their structure constants have been  deduced by ~{\discf, \discs}
 by using symmetry arguments.
One has
\eqn\grav{\eqalign{
U_{l_1|m_1}(z)U_{l_2|m_2}(w)\sim{(z-w)^{-1}}
{C(l_1,l_2,l_3|m_1,m_2,m_3)}f(l_1,l_2)U_{l_3,m_3}
}}
where the $SU(2)$ Clebsch-Gordan coefficients are fixed by the symmetry while
the function of Casimir eigenvalues $f(l_1,l_2)$ is nontrivial 
and was deduced to be given by ~{\discf, \discs}
\eqn\grav{\eqalign{
f(l_1,l_2)
={{{\sqrt{l_1+l_2}}(2l_1+2l_2-2)!}\over{{\sqrt{2l_1l_2}}(2l_1-1)!(2l_2-1)!}}
}}
Remarkably, these structure constants coincide (up to a simple
field redefinition) exactly
with those of $w_\infty$ wedge, defining the asymptotic symmetries of the 
higher spin algebra in $AdS_3$ in a certain basis, computed in  a rather
different context ~{\soojong}.
Thus the primaries (15) are connected to a vertex operator realization of
$AdS_3$ higher-spin algebra.
The related OSFT solution is then constructed similarly to the previous one. It
is given simply by 
\eqn\grav{\eqalign{
\Psi=\sum_{l,m}\lambda^{l|m}{{U}}_{l|m}}}
with the constants 
$\lambda^{l|m}$ satisfying the $\beta$-function 
condition
\eqn\grav{\eqalign{S_{l_1m_1|l_2m_2}^{l_3|m_3}\lambda^{l_1m_1}\lambda^{l_2m_2}=0}}
where
$S_{l_1m_1|l_2m_2}^{l_3|m_3}=C(l_1,l_2,l_3|m_1,m_2,m_3)f(l_1,l_2)$
are the $AdS_3$ higher spin algebra's structure constants.
As previously, this solution describes the perturbative background's
change
\eqn\grav{\eqalign{Q\rightarrow{\tilde{Q}}=Q+\sum_{l,m}\lambda^{l|m}{{U}}_{l|m}}}.
The higher order contributions to the $\beta$-function then will appear
in the SFT solutions with $Q$ replaced with ${\tilde{Q}}$ etc.
Our particular 
goal is, roughly speaking, to find the off-shell analogues of the string field (18)
solving the SFT equation of motion (1),
with $\lambda$-constants satisfying the constraints related to the structures
of the higher-spin algebras.

For that, it is first instructive to investigate  the manifest form of the
operators (note that in ~{\discf}, {\discs}) the structure constants were computed
from the symmetry arguments, without pointing out the explicit form of 
the operators).
Taking the highest weight vector $V_l$  (14) and applying $T_{-}$ using
the OPE
\eqn\grav{\eqalign{ e^{-{i}X{\sqrt{2}}}(z)
e^{(ilX+(l-1)\varphi){\sqrt{2}}}(w)=\sum_{k=0}^\infty
(z-w)^{k-2l}B^{(k)}_{-i{\sqrt{2}}X}:e^{(i(l-1)X+(l-1)\varphi){\sqrt{2}}}:(w)
}}
we obtain
\eqn\grav{\eqalign{
U_{l|l-1}=T_{-}V_l=\oint{dw}:B^{(2l-1)}_{-i{\sqrt{2}}X}e^{(i(l-1)X+(l-1)\varphi){\sqrt{2}}}:(w)
}}
Here 
$
B^{(n)}_{f(z)}\equiv{B^{(n)}}(\partial_z{f},...,\partial_z^n{f})$
are the rank $n$ normalized Bell polynomials in the derivatives of
$f$, defined according to
\eqn\grav{\eqalign{{B^{(n)}}(\partial_z{f},...,\partial_z^n{f})
=
{B^{(n)}}(x_1,...x_n)|_{x_k\equiv{\partial^k}f;1\leq{k}\leq{n}}
\cr
=
\sum_{k=1}^n{B_{n|k}}(x_1,...x_{n-k+1})}}
where $B_{n|k}(x_1,...x_{n-k+1})$ are the normalized partial Bell polynomials
defined according to
\eqn\grav{\eqalign{B_{n|k}(x_1,...x_{n-k+1})=
\sum_{p_1,...p_{n-k+1}}{{1}\over{p_1!...p_{n-k+1}!}}x_1^{p_1}({{x_2}\over{2!}})^{p_2}
...({{x_{n-k+1}}\over{(n-k+1)!}})^{p_{n-k+1}}}}
with the sum taking over all the combinations of non-negative  $p_j$ satisfying
\eqn\grav{\eqalign{
\sum_{j=1}^{n-k+1}p_j=k\cr
\sum_{j=1}^{n-k+1}jp_j=n}}
(note that the standard Bell polynomials $P^{(n)}$ are related to the normalized
ones as $P^{(n)}=n!B^{(n)}$; similarly for the partial Bell polynomials)
To calculate 
the next vertex operator,
$U_{l|l-2}=T_{-}U_{l|l-1}$ one needs to point out, apart from the OPE (21), 
the OPE between 
Bell polynomials of the $X$-derivatives and the exponents 
of $X$ as well.
Using the definitions (23)-(25), it is straightforward to deduce the identity

\eqn\grav{\eqalign{B^{(n)}_{\alpha{X}}(z)e^{\beta{X}}(w)=
\sum_{k=0}^n(z-w)^{-k}{{\Gamma(-\alpha\beta+1)}\over{k!\Gamma(-\alpha\beta+1-k)}}
:B^{(n-k)}_{\alpha{X}}(z)e^{\beta{X}}(w):}}
where $\alpha$ and $\beta$ are some numbers and $\Gamma$ is the Euler's
gamma-function.
Note that this is the double point OPE (sufficient for our purposes),
i.e. accounting only for the contractions between $B^{(n)}_{\alpha{X}}$
and $e^{\beta{X}}(w)$, but not for the expansions
of any of them around some fixed point (such as $z$, $w$ or a midpoint).

Using (21), (26), it is then straightforward to obtain:
\eqn\grav{\eqalign{U_{l|l-2}=2!\oint{dw}
:e^{(i(l-2)X+(l-1)\varphi){\sqrt{2}}}(
B^{(2l-1)}_{-i{\sqrt{2}}X}B^{(2l-3)}_{-i{\sqrt{2}}X}
-(B^{(2l-2)}_{-i{\sqrt{2}}X})^2):(w)}}
This  operator is given by the exponent 
multiplied by the quadratic combination of the Bell polynomials
with ranks $B^{(2l-k_j)};j=1,2$ with $k_1+k_2$ being the length 2
partition of $2^2=4$ with $1\leq{k_{1,2}}\leq{2\times{2}-1=3}$.
It is straightforward to continue this sequence
of transformations by $T_{-}$
to identify the 
manifest expressions for all the vertex operators.
For arbitrary $U_{l|l-m}$ ($1\leq{m}\leq{l}$) we obtain
\eqn\grav{\eqalign{U_{l|l-m}=
m!\oint{dw}e^{(i(l-m)X+(l-1)\varphi){\sqrt{2}}}
\sum_{m^2|k_1...k_m}(-1)^{\pi(k_1,...,k_m)}B_{-i{\sqrt{2}}X}^{(2l-k_1)}
B_{-i{\sqrt{2}}X}^{(2l-k_2)}...B_{-i{\sqrt{2}}X}^{(2l-k_m)}}}
with the sum taken over all the ordered length $m$ partitions of
$m^2=k_1+...+k_m$ with $1\leq{k_1}\leq....\leq{k_m}\leq{2m-1}$
and with the parity  $\pi(k_1,...,k_m)$ of each partition defined as follows.
Consider a particular partition of $m^2$:
$k_1\leq{k_2}....\leq{k_m}$. By permutation we shall call any exchange 
between two neighbouring elements of the partition with one unit that
does not break the order of the partition, e.g.
\eqn\grav{\eqalign{
{\lbrace}k_1\leq{...}{k_{i-1}}{\leq}k_i\leq{k_{i+1}}\leq{k_{i+2}}\leq...\leq{k_m}
{\rbrace}
\cr
\rightarrow
{\lbrace}k_1\leq{...}{k_{i-1}}{\leq}(k_i\pm{1})
\leq({k_{i+1}}\mp{1})\leq{k_{i+2}}\leq...\leq{k_m}
{\rbrace}
}}
Then $\pi(k_1,...,k_m)$  for any length $m$ partition of $m^2$
is the minimum number of permutations
needed to obtain the partition $m^2=k_1+....+k_m$
from the reference partition $m^2=1+3+5+...+(2m-1)$.
Note that, possibly up to an overall sign change of  $U_{l|m}$, any partition 
can be chosen as a reference partition.
One particular lesson that we learn from (28) is that
combinations of the objects of the type $\sum_{\lbrace{n_1}...
{n_k}\rbrace;N=n_1+...+n_k}\alpha_{n_1...n_k}\prod_{j=1}^k{B^{(n_j)}}$
form a basis for the operator realization of the higher-spin algebra.
As we will see below, this is not incidental, as
the products of the Bell polynomials
naturally realize $w_\infty$ and envelopings of $SU(2)$.
In general, they are not primary fields, except for
some very special choices of the $\alpha_{n_1...n_k}$ coefficients
in the summation over the partitions (18).
(strictly speaking, the products of Bell polynomials in (18), (19)
 are the primaries
only for $m=l$; otherwise they must be dressed with the exponents).
Two important numbers characterizing these objects are $N$ and $k$
(total conformal dimension and the partition length).

The ansatz for the solution in $D$-dimensional
string field theory that we propose is the following.
Define the generating function for the Bell polynomials:
\eqn\grav{\eqalign{H(B)=\sum_{n=1}^\infty
{h_n}B^{(n)}_{{\vec{\alpha}}{\vec{X}}}\equiv
\sum_{n=1}^\infty
{{{h_n}P^{(n)}_{{\vec{\alpha}}{\vec{X}}}}\over{n!}}
}}
where 
$P^{(n)}$ are the standard (non-normalized) Bell polynomials
in the derivatives of ${{\vec{\alpha}}{\vec{X}}}$,
$h_n$ are some coefficients, defining the
 associate characteristic function
\eqn\lowen{H(x)=\sum_n{{h_nx^n}\over{n!}}}
This function is convenient to use in order to 
perform various operations with $H(B)$;
e.g. the derivative function $H^\prime(B)$ can be obtained
by differentiating $H(x)$ over $x$ and then replacing
${{x^n}\over{n!}}\rightarrow{B^{(n)}_{{\vec{\alpha}}{\vec{X}}}}$
in the expansion series obtained by differentiation.
Next, define another characteristic function
\eqn\grav{\eqalign{
G(x)=\sum_{n=0}^\infty{{g_nx^n}\over{n!}}}}
Then the composite function $G(H(B))$ generates the products of
Bell polynomial operators according to the Faa de Bruno formula
(which is easy to check by simple straightforward computation):
\eqn\grav{\eqalign{G(H(B))=\sum_{n=0}^{infty}{{g^n}\over{n!}}
\sum_{N=n}^\infty{N!}\sum_{N|k_1...k_n}
h_{k_1}....h_{k_n}
{B^{(k_1)}_{{\vec{\alpha}}{\vec{X}}}}
...{B^{(k_n)}_{{\vec{\alpha}}{\vec{X}}}}\sigma^{-1}(k_1,...k_n)
\cr
=\sum_{n=0}^{\infty}{{g^n}\over{n!}}\sum_{N=n}^\infty
B_{N|n}(h_1{B^{(1)}_{{\vec{\alpha}}{\vec{X}}}},...,
h_{N-n+1}{B^{{(N-n+1)}_{{\vec{\alpha}}{\vec{X}}}}})}}
where $\sum_{N|k_1...k_n}$ stands for the summation 
over ordered length $n$ partitions of $N$
$(0<k_1\leq{k_2}...\leq{k_n})$
and the sigma-factor
\eqn\grav{\eqalign{
\sigma(k_1,...,k_n)=q_{k_1}!...q_{k_n}!}}
is the product of the multiplicities of the elements $k_j$ of the partition.
( note that each $q_{k_j}$ elements $k_j$ entering the partition give rise to
the single factor of $q_{k_j}!$ in the $\sigma^{-1}$ denominator in (33))
We  will be looking for the ansatz SFT solution in the form (35),
that is,
\eqn\grav{\eqalign{\Psi=G(H(B))}}
and our goal is to determine the coefficients
$h_n$ and $g_n$ (more precisely, the defining constraints
on these coefficients imposed by the SFT equations of motion).
${\vec\alpha}$ is some parameter which a priori 
is not fixed; however, we shall see below 
that  the star product for $\Psi$ is drastically simplified
if the tachyon-like constraint: $\alpha^2=-2$ is imposed on 
${\vec{\alpha}}$ and 
it is precisely this simplification that ultimately makes it possible
to formulate the SFT solutions in terms of the functional relations
for $G(H)$.
As it is clear from  (33)-(35), the SFT ansatz that we propose is 
 given by the
series in the partial Bell polynomials of the Bell polynomials
in the target space fields. An essential property of these objects
is that their operator algebra realizes the enveloping of $SU(2)$
with the enveloping parameter related to $\alpha^2$.
In particular, a simple pole in the OPE of these objects
leads to classical $w_\infty$ algebra of area-preserving diffeomorphisms,
while the complete OPE generates the full enveloping ( the
explicit OPE structure will be given below).
This is where the connection with the higher spin algebra enters the game.
To prepare for the analysis of (1) using the ansatz (33)-(35),
 in the next section
we shall analyze the conformal transformation
 properties, operator products and the correlators of the vertex
operators  involving the Bell polynomials and their products.

\centerline{\bf 3. CFT Properties of SFT ansatz}

The first important building block in our construction of the SFT solution
is the analysis of the conformal field theory properties
of operators constructed out of products of the Bell polynomials
in the target space fields. The first step is to determine the transformation
laws for the operators in the sum (33).
We start from the infinitezimal transformation of a single Bell polynomial.
First, we need to evaluate the operator product of the stress tensor with 
$B^{(n)}_{{\vec\alpha}{\vec{X}}}$. This can be done by using
the identity 
\eqn\lowen{\partial_z^n{e^{{\vec{\alpha}}{\vec{X}}}}(z)
=n!B^{(n)}_{{\vec\alpha}{\vec{X}}}n{e^{{\vec{\alpha}}{\vec{X}}}}
}
Then one can deduce the infinitezimal conformal transformation
of $B^{(n)}_{{\vec\alpha}{\vec{X}}}$  
with the generator 
\eqn\lowen{\oint{dz}\epsilon(z)T(z)=-{1\over2}\oint{dz}\epsilon(z)\partial{X_m}
\partial{X^m}(z)}
by using the identity
\eqn\grav{\eqalign{
\delta_\epsilon(\partial_z^n{e^{{\vec{\alpha}}{\vec{X}}}})
=n!(\delta_\epsilon{B^{(n)}_{{\vec\alpha}{\vec{X}}}})e^{{\vec{\alpha}}{\vec{X}}}
+n!{B^{(n)}_{{\vec\alpha}{\vec{X}}}}(\delta_\epsilon{e^{{\vec{\alpha}}{\vec{X}}}})
+\delta_\epsilon(overlap)}}
with $\delta_\epsilon(overlap)$ accounting for the contribution
in which one of the $\partial{X}$'s of $T$ is contracted with $B^{(n)}$
and another with the exponent.
Using the manifest expression for $B^{(n)}_\psi$:
\eqn\grav{\eqalign{
B^{(n)}_\psi=\sum_{l}\sum_{n|p_1...p_l}{{(\partial^{p_1}\psi)^{m_1}...
(\partial^{p_l}\psi)^{m_l}}\over{p_1!...p_k!m_1!...m_k!}}}}
where $\psi={{\vec{\alpha}}{\vec{X}}}$ and
the sum is taken over the ordered partitions
\eqn\grav{\eqalign{
n=\sum_{j=1}^lm_jp_j\cr
k=\sum_{j=1}^l{m_j}\cr
1\leq{k}\leq{n};1{\leq}l\leq{k}\cr
p_1<p_2<...<p_l}}
it is straightdorward to establish the OPE:
\eqn\grav{\eqalign{\partial{X_m}(z)B^{(n)}_\psi(w)
=-\alpha_m\sum_{k=1}^n(z-w)^{-k-1}B^{(n-k)}_\psi(w)
+regular}}
Using the OPE (41)  it is straightforward to compute the overlap
transformation and to deduce the OPE between the stress-energy
tensor and $B^{(n)}_\psi$  with the result given by
\eqn\grav{\eqalign{
T(z)B^{(n)}_\psi(w)=(z-w)^{-1}\partial{B^{(n)}_\psi(w)}
+n(z-w)^{-2}B^{(n)}_\psi(w)+\cr
+\sum_{k=2}^{n+1}(z-w)^{-k-1}(n+1+\alpha^2-{1\over2}(\alpha^2+2)k)
B^{(n-k+1)}_\psi(w)}}
Note that the coefficients  in front of $B^{(n-k+1)}_\psi(w)$
do not depend on $k$ when ${\vec{\alpha}}$ satisfies
the tachyon-like condition $\alpha^2=-2$. This drastically simplifies
the problem to determine the behaviour of $B^{(n)}$ under the global
conformal transformations in the SFT equations, which infinitezimal
form is defined by the OPE (42).
Using the OPE's (41), (42), it is now straightforward to deduce the 
OPE of $T(z)$ with the product of 
 any number $q$ of the Bell polynomials of the target space fields,
which will be the main building block for the SFT solutions that we are 
looking for. 
It is convenient to introduce the notation:
\eqn\grav{\eqalign{R_N^{n_1...n_q}=\prod_{j=1}^qB^{(n_j)}_\psi(w)\cr
N=\sum_j{n_j}}}
We have:
\eqn\grav{\eqalign{T(z)R_N^{n_1...n_q}(w)=
\sum_{j=1}^q\sum_{k_j=2}^{n_j+1}(z-w)^{-k_j-1}
(n_j+1+\alpha^2-{1\over2}(\alpha^2+2)k_j)R_{N-n_j}^{{n_1...n_q}|_{j}}
B_\psi^{(n_j-k_j+1)}\cr
-
\alpha^2\sum_{l,m=1;l<m}^{q}(z-w)^{-k_l-k_m}
(R_{N-n_l-n_m}^{{n_1...n_q}|_{l,m}})
B_\psi^{(n_l-k_l+1)}B_\psi^{(n_m-k_m+1)}
\cr
+N(z-w)^{-2}R_N^{n_1...n_q}(w)+(z-w)^{-1}{\partial}R_N^{n_1...n_q}(w)
}}
where ${{n_1...n_q}|_{j}}$ stands for the set of $q-1$ indices
with $n_j$ excluded, similarly for ${{n_1...n_q}|_{l,m}}$.
Given  the OPE (44)
it is straightforward to obtain the infinitezimal the transformation law for
${cR_N^{n_1...n_q}}$:

We have:
\eqn\grav{\eqalign{{\delta_\epsilon}(cR_N^{n_1...n_q})(w)=
\sum_{j=1}^q\sum_{k_j=2}^{n_j+1}{{\partial^{k_j}\epsilon}\over{k_j!}}
(n_j+1+\alpha^2-{1\over2}(\alpha^2+2)k_j)cR_{N-n_j}^{{n_1...n_q}|_{j}}
B_\psi^{(n_j-k_j+1)}\cr
-
\alpha^2\sum_{l,m=1;l<m}^{q}{{\partial^{k_l+k_m-1}\epsilon}\over{(k_l+k_m-1)!}}
cR_{N-n_l-n_m}^{{n_1...n_q}|_{l,m}}
B_\psi^{(n_l-k_l+1)}B_\psi^{(n_m-k_m+1)}
\cr
+(N-1){\partial\epsilon}R_N^{n_1...n_q}(w)+\epsilon{\partial}(cR_N^{n_1...n_q})(w)
}}

Now we have to establish transformation law for
${cR_N^{n_1...n_q}}$ under $z\rightarrow{f(z)}$, necessary to compute
the correlators in the string field theory equations of motion.
This can be deduced from two conditions:
first, the global transformation should reproduce (45) for
$f(z)=z+\epsilon$. Second, the form of the global transformation  must be 
preserved under the composition. As it is well-known, in case of
simplest non-primary field, such as  the stress-energy tensor,
this leads to the appearance of the Schwarzian derivative of $f(z)$ which is
in fact the degree 2 Bell polynomial in $log(f^\prime(z))$:
\eqn\lowen{S(f(z))=2B_2(-{1\over2}log(f^\prime))}
This is not  a coincidence since for a large class of
non-primaries in $CFT$ the higher degree Bell polynomials 
correspond to the higher derivative extensions of the Schwarzian derivative
in conformal transformations.
Note that the Bell polynomials of the logarithms of functions
defining global conformal transformation satisfy the following composition
identity:
\eqn\grav{\eqalign{B^{(n)}(log({d\over{dx}}f(g(x))))=\sum_{k}B^{(n-k)}(
log(f^\prime(g)))
B^{(k)}(log(g^\prime(x)))}}
making them natural objects present in global conformal transformations.
The global conformal transformations of the Bell polynomials, consistent
with the infinitezimal transformations (45) are deduced to be given by
\eqn\grav{\eqalign{
cR_N^{n_1...n_q}(z)\rightarrow_{z\rightarrow{f(z)}}({{df}\over{dz}})^{N-1}
cR_N^{n_1...n_q}(f(z))
\cr
-
\sum_{j=1}\sum_{k_j=2}^{n_j+1}{1\over{k_j}}
({{df}\over{dz}})^{N-k_j}B^{(k_j-1)}(-(n_j+1+\alpha^2-{1\over2}(\alpha^2+2)k_j)
log({{df}\over{dz}}))
\cr\times
cR_{N-n_j}^{{n_1...n_q}|_{j}}
B_\psi^{(n_j-k_j+1)}\cr
+
\sum_{l,m=1;l<m}^{q}{1\over{(k_l+k_m-1)}}
({{df}\over{dz}})^{N-k_l-k_m+1}B_{k_l+k_m-2}(-\alpha^2log({{df}\over{dz}}))
\cr\times
cR_{N-n_l-n_m}^{{n_1...n_q}|_{l,m}}
B_\psi^{(n_l-k_l+1)}B_\psi^{(n_m-k_m+1)}
}}

For our purposes, we shall need to  compute the
 values of the Bell polynomials  in the transformation law (48) for the
functions $I(z)=-{1\over{z}}$ (in the kinetic term of the SFT action)
and $g\circ{f_{k}^{3}}(z)$ at $z=0$ where $f_k^3$ defined in (5)
map the string worldsheets of the cubic theory to the wedges of the disc
and 
\eqn\lowen{g(z)=i{{1-z}\over{1+z}}}
further maps  this disc to the half-plane, so that
\eqn\grav{\eqalign{g\circ{f_{1}^{3}}(0)=0\cr
g\circ{f_{2}^{3}}(0)={\sqrt{3}}\cr
g\circ{f_{3}^{3}}(0)=-{\sqrt{3}}}}
For that, we shall use the fact that if
$\lbrace{a_n}\rbrace$ are the coefficients in the series
expansion of any function $f(x)=\sum_{n=1}{{a_nx^n}\over{n!}}$
(assume $f(0)=0$), then 
$e^{f(x)}=1+\sum_{n=1}^\infty{B^{(n)}(a_1,...,a_n)}x^n$.
From now on, to abbreviate things, we shall restrict ourselves to the
case $\alpha^2=-2$, relevant to our SFT solution.
We start from $B^{(k)}(\kappa{log(I^\prime(z))})$
where, in particular, $\kappa=1-n_j$ in the first group of terms in (48) and
$\kappa=-\alpha^2=2$ in the second.
Then
\eqn\grav{\eqalign{B^{(n)}(\kappa{log(I^\prime(z))})
=z^{-n}B^{(n)}(2\kappa,-2\kappa,...(-1)^n2(n-1)!\kappa)}}
The Bell polynomial on the right-hand side is then identified with the
$n$'th expansion coefficient of the exponent of $-2\kappa{log}z$,
i.e. of $z^{-2\kappa}$
Therefore
\eqn\grav{\eqalign{B^{(n)}(\kappa{log(I^\prime(z))})
={{\Gamma(1-2\kappa)z^{-n}}\over{n!\Gamma(1-2\kappa-n)}}}}
Next, we need the values of the  Bell polynomials 
$B^{(n)}(log{{{df(z)}\over{dz}}})$ with
$f(z)=g\circ{f_{k}^{3}}(z)$ at $z=0$.
Straightforward calculation gives the result
\eqn\lowen{B^{(n)}({\kappa}log{{{df(z)}\over{dz}}})|_{z=0}=
B^{(n)}(\beta_1...\beta_k...\beta_n)}
with

\eqn\grav{\eqalign{
\beta_k=\kappa({5\over3}(-i)^n+{1\over3}(i)^n-({1\over2})^{n-1})(n-1)!}}

for $g\circ{f_{1}^{3}}(z)$,

\eqn\grav{\eqalign{
\beta_k=\kappa({5\over3}(-i)^n+{1\over3}(i)^n+2e^{{2i\pi{n}}\over{3}})(n-1)!}}

for $g\circ{f_{2}^{3}}(z)$ and

\eqn\grav{\eqalign{
\beta_k=\kappa({5\over3}(-i)^n+{1\over3}(i)^n+2e^{-{2i\pi{n}}\over{3}})(n-1)!}}

for $g\circ{f_{3}^{3}}(z)$.
Accordingly, these Bell polynomials are identified with the expansion
series of
\eqn\grav{\eqalign{
h_1(z)=(1+iz)^{-{5\over3}\kappa}(1-iz)^{-{1\over3}\kappa}(1+{z\over{2}})^{-2\kappa}\cr
h_2(z)=(1+iz)^{-{5\over3}\kappa}(1-iz)^{-{1\over3}\kappa}
(1-e^{{2i\pi}\over3}z)^{-2\kappa}\cr
h_3(z)=
(1+iz)^{-{5\over3}\kappa}(1-iz)^{-{1\over3}\kappa}
(1-e^{-{2i\pi}\over3}z)^{-2\kappa}}}
for
 $g\circ{f_{1}^{3}}$,  $g\circ{f_{2}^{3}}$ and  $g\circ{f_{3}^{3}}$
respectively.
Accordingly, the values of the Bell polynomials are given by
\eqn\grav{\eqalign{
B^{(n)}(\kappa{log}({d\over{dz}}g\circ{f_{1}^{3}}(z)))|_{z=0}
\cr
=\sum_{k,l,m|k+l+m=n}{{e^{{{i\pi}\over2}(k-l)}2^{-m}}\over{k!l!m!}}
{{\Gamma(1-{5\over3}\kappa)\Gamma(1-{1\over3}\kappa)\Gamma(1-2\kappa)}
\over{\Gamma(1-{5\over3}\kappa-k)\Gamma(1-{1\over3}\kappa-l)\Gamma(1-2\kappa-m)}
}
\cr
B^{(n)}(\kappa{log}({d\over{dz}}g\circ{f_{2}^{3}}(z)))|_{z=0}
\cr
=\sum_{k,l,m|k+l+m=n}{{e^{{i\pi}({1\over2}(k-l)+{{2m}\over3})}}\over{k!l!m!}}
{{\Gamma(1-{5\over3}\kappa)\Gamma(1-{1\over3}\kappa)\Gamma(1-2\kappa)}
\over{\Gamma(1-{5\over3}\kappa-k)\Gamma(1-{1\over3}\kappa-l)\Gamma(1-2\kappa-m)}
}
\cr
B^{(n)}(\kappa{log}({d\over{dz}}g\circ{f_{3}^{3}}(z)))|_{z=0}
\cr
=\sum_{k,l,m|k+l+m=n}{{e^{{i\pi}({1\over2}(k-l)-{{2m}\over3})}}\over{k!l!m!}}
{{\Gamma(1-{5\over3}\kappa)\Gamma(1-{1\over3}\kappa)\Gamma(1-2\kappa)}
\over{\Gamma(1-{5\over3}\kappa-k)\Gamma(1-{1\over3}\kappa-l)\Gamma(1-2\kappa-m)}
}}}
with the sums taken over
the unordered partitions of $n=k+l+m$.
These relations altogether fully determine  the transformation properties
of our string field ansatz, including the star product.

The final step to make before actually computing the SFT correlators 
is to point out the operator product rules involving the Bell polynomial
operators and their blocks. We will do this in the next section, in particular
deriving an analogue of the  generalized Wick's theorem for the Bell
polynomial operators and pointing the relevance of their correlators
to the structure constants of the higher-spin algebra.

\centerline{\bf 4. Bell Polynomial Operators: Operator Products and 
Correlators}

The most crucial building block in our computations involves the OPE rules
for the operators of the SFT ansatz (33)-(35)
 which we will establish in this section.
Ultimately, it turns out that 
it is precisely the structure of these OPE rules which makes 
it possible to work out the SFT solution and, moreover, 
to relate it to the higher spin algebra.

We start from the simplest OPE between
$B_{{\vec\alpha}{\vec{X}}}^{(N)}(z)$ and$ B_{{\vec\beta}{\vec{X}}}^{(m)}(w)$.
This  doesn't turn out to be an easy OPE to compute .
The manifest expressions
(23)-(24)
 for the Bell polynomials do not appear to be very helpful. Nevertheless, 
there are some observations to simplify the computation.
First of all, the OPE has to preserve the conformal transformation structure
(48) of the Bell polynomial operators.
This suggests that the OPE must have the structure
\eqn\grav{\eqalign{B_{{\vec\alpha}{\vec{X}}}^{(N)}(z)B_{{\vec\beta}{\vec{X}}}^{(M)}(w)
=\sum_{n=0}^{N}\sum_{m=0}^M(z-w)^{-n-m}\lambda^{N|M}|_{m|n}
:B_{{\vec\alpha}{\vec{X}}}^{(N-n)}(z)B_{{\vec\beta}{\vec{X}}}^{(M-m)}(w):}}
(again, for the brevity we consider the double point OPE here, just as 
was explained above). In other words, the Bell polynomial structure (48)
 of the 
operators is preserved by (59).
The next helpful hint comes from the identity (36) 
relating the Bell operators  to the derivatives of the exponents
and from analyzing the correlator
\eqn\grav{\eqalign{
<B_{{\vec\alpha}{\vec{X}}}^{(N)}e^{{\vec\alpha}{\vec{X}}}
(z)B_{{\vec\beta}{\vec{X}}}^{(M)}e^{{\vec\beta}{\vec{X}}}(w)\cr
={1\over{N!M!}}\partial_z^N\partial_w^M<
e^{{\vec\alpha}{\vec{X}}}(z)e^{{\vec\beta}{\vec{X}}}(w)>
=(z-w)^{-{\vec\alpha}{\vec\beta}-N-M}{{\Gamma(1-{\vec\alpha}{\vec\beta})}\over
{N!M!\Gamma(1-{\vec\alpha}{\vec\beta}-M-N)}}}}
This correlator can be computed in two equivalent ways:
one either starts with applying the OPE (26) 
of the Bell polynomials with the exponents
and then contracting the remaining  derivatives of $X$ 
between themselves in each of the OPE terms -
or, alternatively, starting with the OPE (59)
 between the Bell polynomials, containing the
unknown $\lambda$-constants and then contracting
the remaining derivatives of X in each of the operators 
with the opposite exponent.
Comparison of these two expressions identifies 
the remarkably simple OPE structure:
\eqn\grav{\eqalign{B_{{\vec\alpha}{\vec{X}}}^{(N)}(z)B_{{\vec\beta}{\vec{X}}}^{(N)}(w)
=<B_{{\vec\alpha}{\vec{X}}}^{(n)}(z)
B_{{\vec\beta}{\vec{X}}}^{(m)}(w)>:B_{{\vec\alpha}{\vec{X}}}^{(N-n)}(z)
B_{{\vec\beta}{\vec{X}}}^{(M-m)}(w):}}
i.e. the OPE coefficients are simply given by the
 two-point correlators of the lower rank polynomials:
\eqn\grav{\eqalign{\lambda^{N|M}|_{m|n}\equiv\lambda_{m|n}
=(z-w)^{n+m}<B_{{\vec\alpha}{\vec{X}}}^{(n)}(z)B_{{\vec\beta}{\vec{X}}}^{(m)}(w)>}}
The last step is to compute the two-point correlators and somehow
 this again doesn't turn out to be an elementary exercise.
 Straightforward calculation using
the manifest expression (24) for the Bell polynomials
and the Wick's theorem leads to complicated sum over 
partitions which doesn't seem to be realistic to evaluate
and doesn't look  illuminating or useful for our purposes. Instead, 
we shall start from the  identity
\eqn\grav{\eqalign{B^{(n)}_{{\vec\alpha}{\vec{X}}}
={1\over{n}}({\partial}B^{(n-1)}_{{\vec\alpha}{\vec{X}}}
+{\vec\alpha}\partial{\vec{X}}B^{(n-1)}_{{\vec\alpha}{\vec{X}}})}}
Inserting this identity in 
the correlator (62) and using the OPE (59) we obtain the recursion 
relation
\eqn\grav{\eqalign{\lambda_{n|m}=
-{{n+m-1}\over{n}}\lambda_{n-1|m}-{{{\vec{\alpha}}{\vec{\beta}}}\over{n}}
\sum_{l=1}^{m-1}\lambda_{n-1|l}}}
This recursion relation can be simplified
 by repeating the above procedure and inserting the identity (63)
into the correlator $<B^{(n)}B^{(m-1)}>$, 
obtaining the similar recursion relation
for $\lambda_{n|m-1}$ and subtracting it from (64). Then the recursion becomes
\eqn\grav{\eqalign{n(\lambda_{n|m}-\lambda_{n|m-1})=-(n+m-1)\lambda_{n-1|m}
+(n+m-2-{\vec\alpha}{\vec\beta})\lambda_{n-1|m-1}}}
with the obvious physical constraints
\eqn\grav{\eqalign{\lambda_{0|k}=\lambda_{k|0}=\delta_{0k}}}
To solve this recursion, define the generating function
\eqn\lowen{F_\lambda(x,y)=\sum_{m,n}\lambda_{n|m}x^ny^m,}
multiply the recursion (65) by $x^ny^m$ and sum over $m$ and $n$.
This leads to the first order partial differential equation 
for $F_\lambda(x,y)$:
\eqn\grav{\eqalign{(1-y)(1+x)\partial_x{F_\lambda}+y(1-y)\partial_y{F_\lambda}
+{\vec{\alpha}}{\vec{\beta}}y{F}=0}}
with the boundary conditions
\eqn\lowen{F_\lambda(x,0)=F_\lambda(0,y)=1}
This equation isn't hard to solve.
Defining
\eqn\grav{\eqalign{\xi=log(1+x),\eta=log(y)\cr
G(x,y)=log{F(x,y)}}}
the equation simplifies according to
\eqn\grav{\eqalign{\partial_\xi{G(\xi,\eta)}+\partial_\eta{G(\xi,\eta)}-
{{{\vec{\alpha}}{\vec{\beta}}y}\over{1-e^{-\eta}}}=0}}
and is equivalent to the characteristic ODE system
\eqn\grav{\eqalign{{{d\xi}\over{ds}}={{d\eta}\over{ds}}=1\cr
{{d{G}}\over{ds}}=
{{{\vec{\alpha}}{\vec{\beta}}y}\over{1-e^{-\eta}}}}}
so the general solution is
\eqn\grav{\eqalign{
G(\xi,\eta)=H(\xi-\eta)+{\vec{\alpha}}
{\vec{\beta}}\int{{d\eta}\over{1-e^{-\eta}}}}}
Substituting  $G(0,\eta)=0$
then fixes $H$ to be
\eqn\grav{\eqalign{H(\xi-\eta)={\vec{\alpha}}{\vec{\beta}}log(1-e^{\eta-\xi})}}
so the solution is
\eqn\grav{\eqalign{
F_\lambda(x,y)=({{(1+x)(1-y)}\over{(1+x-y)}})^{-{\vec{\alpha}}{\vec{\beta}}}\cr
\lambda_{n|m}={1\over{n!m!}}\partial^n_x\partial^m_y{F_\lambda(x,y)|_{x,y=0}}}}
This solution, 
describing the correlator of two Bell polynomial operators
is related to the higher-spin algebra in $AdS_3$ 
and determines the parameter $\mu$ of the enveloping
$T(\mu)$ of $SU(2)$ ~{\discf, \discs, \romans}.

Next, using the OPE (64), (75), 
it is straightforward to identify the worldsheet correlators 
of the products of the Bell  operators  in terms of the $\lambda_{m|n}$
-numbers, 
 relevant to $w_\infty$ and to the SU(2) enveloping
generators, as well as to our SFT ansatz (33)-(35).
The result is given by
\eqn\grav{\eqalign{
<:B^{(n_1)}_{\vec{\alpha}}...B^{(n_{p})}:(z):B^{(m_1)}
...B^{(m_q)}:(w)>|_{N=n_1+...+n_p;M=m_1+...+m_q}
\cr
=(z-w)^{-N-M}
\sum_{partitions{\lbrack}\alpha_{ij},\beta_{ji}{\rbrack}}\prod_{i=1}^p\prod_{j=1}^q
(q!)^p\lambda_{\alpha_{ij}|\beta_{ji}}\cr
(\prod_{k=1}^{q}\sigma(\alpha_{1k}|\beta_{k1})!)^{-1}...
(\prod_{k=1}^q\sigma(\alpha_{pk}|\beta_{kp})!)^{-1}
(\prod_{l=1}^p\sigma(s_l)!)^{-1}}}
with the constraints
\eqn\grav{\eqalign{
\sum_{j=1}^{q}\alpha_{ij}=n_i\cr
\sum_{i=1}^{q}\beta_{ji}=m_j\cr
\sum_{i=1}^p\alpha_{ij}=r_j\cr
\sum_{j=1}^q\beta_{ji}=s_i\cr
\sum_{j=1}^q{r_j}=\sum_{i=1}^p{n_i}=N\cr
\sum_{j=1}^q{m_j}=\sum_{i=1}^p{s_i}=N\cr
}}
where the notations are as follows.
We have introduced the {\it exchange numbers}
$\alpha_{ij}$ indicating how much of the total  conformal dimension 
 $n_i$ of
the $B^{(n_i)}(z)$-operator in the product of the Bell polynomials
on the left at $z$ is contributed  to its  interaction 
with the operator $B^{(m_j)}(w)$ in the product of the Bell polynomials
at $w$ on the right, according to the OPE structure (59).
 Similarly, the exchange number $\beta_{ji}$ indicates 
the reduction in the conformal dimension of  $B^{(m_j)}(w)$ on the right
as a result of its interaction to $B^{(n_i)}(z)$ on the left.
Altogether, this corresponds to the order of $(z-w)^{-\alpha_{ij}-\beta_{ji}}$
term in the OPE of these two operators entering the left and the right chains,
contributing to the overall correlator.
Thus $r_j$-numbers, forming the length $q$ partition of $N$
(as opposed to the length $p$ partition of $N$, formed by $n_i$)
indicate the total loss of  conformal dimension of 
the complete operator on the left-hand side at $z$
due to the interaction with the single polynomial
$B^{(m_j)}(w)$ on the right.
Similarly, the $s_i$-numbers, forming the  length $p$ partition of $M$
(as opposed to the length $q$ partition of $M$ formed by $m_i$),
indicate the total loss of  conformal dimension of
the complete operator on the right-hand side
at $w$  due to the interaction with the single polynomial
$B^{(n_i)}(z)$ on the left.
Next, $\sigma(\alpha_{jk}|\beta_{kj})(j=1,...,p) $
indicates the multiplicity of the array of the exchange numbers
$\alpha_{jk}|\beta_{kj}$ in $p$ arrays of the length $q$ each:
${\lbrace}\alpha_{j1}|\beta_{1j},...,{\alpha_{jq}|\beta_{qj}}{\rbrace}$
($j=1,...p$), similarly to (34).
Finally,
$\sigma(s_l)$
counts multiplicities of the $s$-numbers defined above
(again, similarly to (34))
As before, all the partitions are considered ordered.
While the sum (76) involving the products of the exchange numbers,
summed over the partitions, looks tedious, there are some significant
simplifications in important cases, when the partitions are summed over.

In fact, we are particularly interested
 in objects of the type (43) with the partitions summed 
over, as, e.g. in (33)-(35).
Let us again start with the simplest possible warm-up example 
of summing over the partitions
- with all the partition elements summed over uniformly,that is,
with the sum being a Bell polynomial of Bell polynomials.
Namely, consider the elementary example of a toy string field given by
\eqn\grav{\eqalign{\Psi=\sum_{M=0}^\infty\sum_{q=1}^M
\sum_{M|m_1...m_q}(\prod_{i=1}^q{\lbrack\sigma(m_i)\rbrack^{-1}})
B^{(m_1)}....B^{(m_q)}}}
and let's calculate the simplest SFT correlator $<\Psi(1)\Psi(0)>$.
To calculate this correlator, the expression (76) 
must be further summed over the partitions
according to the definition (78) of the toy $\Psi$.
Take the the product (76), defining the
 correlator $<B^{(n_1)}....B^{(n_p)}(1)B^{(m_1)}....B^{(m_q)}(0)>$
and let us begin with the summation 
over  partitions in the second operator at $0$.
Consider the first row in the product (76)
\eqn\grav{\eqalign{\sim\sum_{partitions(n_1,s_1)}\prod_{j=1}^q
(q!)\lambda_{\alpha_{ij}|\beta_{ji}}
(\prod_{k=1}^{q}\sigma(\alpha_{1k}|\beta_{k1}))^{-1}
}}
with the sum taken over the partitions of $n_1$ and $s_1$ 
into the exchange number sets.
This row completely describes the interaction of $B^{(n_1)}$ 
with the array of the Bell polynomials
at $w$ with $M$ being the total conformal dimension of the array.
Let us calculate the effect of the partition summation (78)  for this row.
Now, in addition to the summation over the above partitions,
 sum over all the partitions of $M$ with lengths $1\leq{q}\leq{M}$ and
uniform weights for each $q$. 

It is then straightforward to check that the result will be given by
the series expansion coefficient of the following simple generating function:

\eqn\grav{\eqalign{\sim\sum_{partitions(n_1,s_1)}\prod_{j=1}^q
(q!)\lambda_{\alpha_{ij}|\beta_{ji}}\cr
(\prod_{k=1}^{q}\sigma(\alpha_{1k}|\beta_{k1})!)^{-1}
\cr
={1\over{n_1!s_1!}}\partial_{x}^{n_1}\partial_y^{s_1}
({1\over{1-F_\lambda(x,y)}})|_{x,y=0}
}}
The same  procedure can be repeated for the remaining $p-1$ rows 
parametrized
by $(n_j,s_j),j=1,...p$, leading to
the 
$$\sim{{\partial_x^{n_1}\partial_y^{s_1}({1\over{1-F_\lambda(x,y)}})...
{\partial_x^{n_p}\partial_y^{s_p}
({1\over{1-F_\lambda(x,y)}})}}\over{n_1!s_1!...n_p!s_p!
\lbrack\sigma(n_1|s_1)\rbrack!...\lbrack\sigma(n_p|s_p)\rbrack!}}$$
Finally, let us sum over the partitions for the first string field at 1.
For the fixed values of $N$ and $M$ the result is 
\eqn\grav{\eqalign{\sim{1\over{M!N!}}
{\partial_x^N\partial_y^M({1\over{1-{1\over{1-F_\lambda}}}})|_{x,y=0}}=
-\partial_x^N\partial_y^M{1\over{F_\lambda}}|_{x,y=0},
}}
so the two-point function of the toy string field (78) is
\eqn\grav{\eqalign{
<\Psi(1)\Psi(0)>=e^{-{1\over{F_\lambda(x,y)}}}|_{x=y=1}}}
The objects of the type (78) are of interest to  us both because
they are relevant to our SFT ansatz and, at the same time, form an
operator algebra realization of $w_\infty$ and $SU(2)$ envelopings.
Namely, instead of the string field $\Psi$ (78)
consider the field
\eqn\grav{\eqalign{\Psi_{N|p}=\sum_{q=1}^p
\sum_{N|n_1...n_q}(\prod_{i=1}^q{\lbrack\sigma(n_i)\rbrack^{-1}})
B^{(n_1)}....B^{(n_q)}}}

(it is easy to see that  the toy string field $\Psi$
is given by $\Psi=\sum_{N=0}^\infty\Psi_{N|N}$)
This field is characterized by the numbers $N$ and $p$, with the first
being its total conformal dimension and the second indicating the maximum
length of the ``words'' made out of
 Bell polynomial ``letters'', contained in the 
string field ``sentence'' $\Psi_{N|p}$. Let us compute the OPE
of two ``sentences'' $\Psi_{N_1|p_1}(z)$ and $\Psi_{N_2|p_2}(w)$
around the midpoint ${1\over2}(z+w)$. Clearly, the conformal transformation
properties of the Bell polynomials imply that the Bell polynomial structure
must be preserved under such an operator product. It is also clear,
from the OPE structure (61), (76) for the Bell polynomials
that the terms (``sentences'')of the order $(z-w)^{-N}(N>0)$ would
consist  of words of conformal dimension $N_1+N_2-N$ and 
lengths up to $p_1+p_2$.
For $N=1$ this sends a strong hint towards the emergence of $w_\infty$ and
of $SU(2)$ envelopings for higher order $N$ values.
Indeed, straightforward calculation, using (61) 
and the recurrence relation (63)
leads to the following midpoint OPE simple pole:
\eqn\grav{\eqalign{\Psi_{N_1|p_1}(z)\Psi_{N_2|p_2}(z)
=(z-w)^{-1}(N_2p_1-N_1p_2)\Psi_{N_1+N_2-1|p_1+p_2}({{z+w}\over2})}}
and the general OPE structure:
\eqn\grav{\eqalign{\Psi_{N_1|p_1}(z)\Psi_{N_2|p_2}(z)
=(z-w)^{-1}(N_2p_1-N_1p_2)\Psi_{N_1+N_2-1|p_1+p_2}({{z+w}\over2})
\cr
+\sum_{n=2}^{N_1+N_2}(z-w)^{-n}\gamma_n(N_1,p_1|N_2,p_2)
\Psi_{N_1+N_2-n|p_1+p_2}({{z+w}\over2})
}}
Although we have not computed the $\gamma_n$ coefficients
in this paper explicitly, such a computation doesn't look a 
conceptual challenge and the result must be
 anyway determined by combinations of the $\lambda$-numbers (75) 
stemming from the two-point correlators of the Bell polynomials.
So we recognize classical $w_\infty$ at the simple pole and 
the enveloping  $T(\mu)$ of SU(2) at
the higher order singularities with the $\mu$-parameter related
to the $\lambda$-numbers.
Note that this is the midpoint OPE. If, for example, one needs to compute
the OPE around the $w$-point, the right-hand side of
(85) must be shifted from ${1\over2}(z+w)$ to $w$ by the appropriate
series expansion in ${1\over2}(z-w)$. This way, the full enveloping
algebra will appear, for example, in  commutators
of the charges $\oint{dz}\Psi_{N|n}$.

Now consider a more general example of a string field, 
relevant to our ansatz.
Consider again a generating function of the normalized Bell polynomials:
$$H(B)=
\sum_{n=0}^\infty{{h_n}\over{n!}}B^{(n)}_{{\vec\alpha}{\vec{X}}}$$
and the string field given by
$$\Psi=G\circ{H(B)}$$
where
$$G(x)=\sum_{n=0}^\infty{{g_n}\over{n!}}x^n$$
with some fixed coefficients $h_n$ and $g_n$.
Using the OPE rules and the formalism developed above, the calculation of 
the two-point function gives:
\eqn\grav{\eqalign{
<\Psi(1)\Psi(0)>=
\cr
\sum_{N=0}^\infty\sum_{M=0}^\infty
\sum_{n=0}^N\sum_{m=0}^M{{g_ng_m}}\sum_{N|p_1...p_n}\sum_{M|q_1...q_m}
{{{h_{p_1}...h_{p_{n}}}{h_{q_1}...h_{q_m}}}\over{p_1!...p_n!q_1!...q_m!
\prod_{i,j}\lbrack\sigma(p_i)\rbrack!
\lbrack\sigma(q_j)\rbrack!}}
\cr\times
\sum_{partitions{\lbrack}\alpha_{ij},\beta_{ji}{\rbrack}}
\prod_{i=1}^n\prod_{j=1}^m\lambda_{\alpha_{ij}|\beta_{ji}}
(\prod_{k=1}^{m}\sigma(\alpha_{1k}|\beta_{k1})!)^{-1}...
(\prod_{k=1}^m\sigma(\alpha_{nk}|\beta_{kn})!)^{-1}
(\prod_{l=1}^n\sigma(s_l)!)^{-1}}}
with the constraints
\eqn\grav{\eqalign{
\sum_{j=1}^{m}\alpha_{ij}=n_i;
\sum_{i=1}^{n}\beta_{ji}=m_j\cr
\sum_{i=1}^n\alpha_{ij}=r_j;
\sum_{j=1}^m\beta_{ji}=s_i\cr
\sum_{j=1}^m{r_j}=\sum_{i=1}^n{n_i}=N\cr
\sum_{j=1}^m{m_j}=\sum_{i=1}^n{s_i}=N
}}
with the multiplicity $\sigma$-factors,
exchange numbers and $r,s$-numbers defined as before.
By direct comparison, it is straightforward to realize that the 
lengthy expression 
on the right-hand side of (86), (87) is just a series expansion  of the
 relatively simple
generating composite function, that is, it
can be cast as simply as
\eqn\grav{\eqalign{<\Psi(1)\Psi(0)>=
{\tilde{G}}(H(F_\lambda(x,y));H(F_\lambda(x,y)))|_{x=y=1}}}
where the function of two variables ${\tilde{G}}(x,y)$ is
related to the function $G(x)$ with the single argument according to
\eqn\grav{\eqalign{{\tilde{G}}(x,y)=\sum_{m,n}{{g_mg_n}\over{m!n!}}x^my^n}}
where $g_n$ are the expansion coefficients of $G$.
It is instructive to generalize this two-point correlator to the case
of two different string fields, that is, for the case of 
string fields of the
 type (84) with the
different $H$-functions, but with the same $G$-function.
The calculation, completely similar to the above, gives:
\eqn\grav{\eqalign{\Psi_1=G\circ{H_1(B)}\cr
\Psi_2=G\circ{H_2(B)}\cr
<\Psi_1(1)\Psi_2(0)>={\tilde{G}}(H_1(F_\lambda(x,y));
H_2(F_\lambda(x,y)))|_{x=y=1}}}
The next step in the  computation  the SFT correlators relevant 
to the equations of motion in SFT
is to determine how the global conformal transformations
by $I(z)$ and $g_k(z)\equiv{g}\circ{f_k^3}(z)$
 act on the string 
fields of the type (84). Using the transformations (48)-(58) , 
it is not difficult
to deduce that, under any of these conformal transformations
(denoted by  $f(z)$ for the brevity)
the string field (84) transforms as
\eqn\grav{\eqalign{\Psi\equiv
G(H(B))\rightarrow{\hat{f}}\Psi
\cr
={{dG}\over{dH}}\sum_{n=1}^\infty\sum_{k=1}^n\alpha_f(k,n)
\lbrack{{{(B-1)}}}\partial^k_B{H(B)}+
\partial_B^{k-1}H(B)\rbrack
\cr
+{{d^2G}\over{dH^2}}\sum_{m,n=1;m<n}^\infty\sum_{k,l=1;m<n}^{k+l=n-1}
\partial^k{H}\partial^l{H}\beta_f(k,l|n)}}
with the differentiation rules for $G$ and $H$ explained above (31), (32) and
with the coefficients $\alpha_f$ and $\beta_f$ 
related to the 
 conformal transformations by $I(z)=-{1\over{z}}$ and  
$g\circ{f_k^3}(z)$ (5),  
(49)
 according to:
\eqn\grav{\eqalign{\alpha_{I(z)}(k,n)={1\over{k!}}
{{\Gamma(2n-1)}\over{\Gamma(2n-k)}}\cr
\alpha_{g\circ{f_1^3}(z)}(k,n)={1\over{k}}(-{2\over3})^{n-k}
\sum_{a,b,c|a+b+c=k-1}{{e^{{{i\pi}\over2}(a-b)}2^{-c}}\over{a!b!c!}}
\cr\times
{{\Gamma(1-{5\over3}(1-n))\Gamma(1-{1\over3}(1-n))\Gamma(1-2(1-n))}
\over{\Gamma(1-{5\over3}(1-n)-a)\Gamma(1-{1\over3}(1-n)-b)\Gamma(1-2(1-n)-c)}}
\cr
\alpha_{g\circ{f_2^3}(z)}(k,n)={1\over{k}}(-{8\over3})^{n-k}
\sum_{a,b,c|a+b+c=k-1}{{e^{({{i\pi}\over2}(a-b)+{{2i\pi{c}}\over3})}2^{-c}}\over{a!b!c!}}
\cr\times
{{\Gamma(1-{5\over3}(1-n))\Gamma(1-{1\over3}(1-n))\Gamma(1-2(1-n))}
\over{\Gamma(1-{5\over3}(1-n)-a)
\Gamma(1-{1\over3}(1-n)-b)\Gamma(1-2(1-n)-c)}}\cr
\alpha_{g\circ{f_3^3}(z)}(k,n)={1\over{k}}({8\over3})^{n-k}
\sum_{a,b,c|a+b+c=k-1}{{e^{({{i\pi}\over2}(a-b)-{{2i\pi{c}}\over3})}2^{-c}}\over{a!b!c!}}
\cr\times
{{\Gamma(1-{5\over3}(1-n))\Gamma(1-{1\over3}(1-n))\Gamma(1-2(1-n))}
\over{\Gamma(1-{5\over3}(1-n)-a)\Gamma(1-{1\over3}(1-n)-b)\Gamma(1-2(1-n)-c)}}
}}
and
\eqn\grav{\eqalign{
\beta_{I(z)}(k,l|n)=
{2\over{(k+l-1)!}}(-{2\over3})^{n-k-l+1}{{\Gamma(-3)}\over{\Gamma(-1-k-l)}}
\cr
\beta_{g\circ{f_1^3}(z)}(k,l|n)
={1\over{k+l-1}}(-{8\over3})^{n-k-l+1}
\cr\times
\sum_{a,b,c|a+b+c=k+l-2}{{e^{{{i\pi}\over2}(a-b)}2^{-c}}\over{a!b!c!}}
{{\Gamma(-{7\over3})\Gamma({1\over3})\Gamma(-3)}
\over{\Gamma(-{7\over3}-a)\Gamma({1\over3}-b)\Gamma(-3-c)}}
\cr
\beta_{g\circ{f_2^3}(z)}(k,l|n)=
{1\over{k+l-1}}(-{8\over3})^{n-k-l+1}
\cr\times
\sum_{a,b,c|a+b+c=k+l-2}{{e^{({{i\pi}\over2}(a-b)+{{2i\pi{c}}\over3})}2^{-c}}\over{a!b!c!}}
{{\Gamma(-{7\over3})\Gamma({1\over3})\Gamma(-3)}
\over{\Gamma(-{7\over3}-a)\Gamma({1\over3}-b)\Gamma(-3-c)}}\cr
\beta_{g\circ{f_2^3}(z)}(k,l|n)=
{1\over{k+l-1}}({8\over3})^{n-k-l+1}
\cr\times
\sum_{a,b,c|a+b+c=k+l-2}{{e^{({{i\pi}\over2}(a-b)-{{2i\pi{c}}\over3})}2^{-c}}\over{a!b!c!}}
{{\Gamma(-{7\over3})\Gamma({1\over3})\Gamma(-3)}
\over{\Gamma(-{7\over3}-a)\Gamma({1\over3}-b)\Gamma(-3-c)}}}}
Furthermore, our notations in (91) are defined as follows.
Consider a function of the normalized 
Bell polynomials $f(B_1,B_2,....)=\sum_{n>{0}}f_nB_{\vec{\psi}}^{(n)}$
and the associate function $f(B)$ given by the formal series 
in auxiliary argument $B$
$f(B)=\sum_{n>0}f_n{B^n}$
Consider a transformation $f(B)\rightarrow{g(B)}$
where $g(B)=\sum_n{g_n}B^n$ can obrained from $f$ by differentiation over $B$,
integration, multiplication(s) by $B$ and/or their combination.
Then the formal series for 
$g(B)$ define the new associate generating function of the normalized
Bell polynomials $g(B_1,...,B_n)=\sum_n{g_n{B_{\vec\psi}^{(n)}}}$ 
by identifying $B^n\rightarrow{B_{\vec\psi}^{(n)}}$.

This fully determines the transformations of the SFT string field ansatz 
under the conformal transformations mapping the 
worlsheets to the wedges of the single
disc and then to the single half-plane.
The next step is to point out the action of the BRST charge on the string 
field ansatz.
This too can be reduced to the transformations of the ansatz functions 
$G$ and $H$.  Since the only SFT correlator involving the BRST charge is 
$<<Q\Psi|\Psi>>=<Q\Psi(0)I\circ\Psi(0)>$ and 
both $\Psi$ and $I\circ{\Psi}$ are proportional to $c$,
the only terms in the commutator with the BRST charge (12) 
contributing to this correlator are those
proportional to $\partial{c}c$  and $\partial^2{c}c$, while all the terms
in $Q\Psi$ containing higher derivatives of the $c$-ghost 
don't contribute to the correlator since
$\partial^n{c}c\sim{B_\sigma^{(n-1)}}\partial{c}c$ with $\sigma$
 being the bosonized $c$-ghost.
Such terms do not contribute to the two-point correlators since 
the Bell polynomials in the derivatives of the bosonized 
 $c$-ghost: $B^{(n-1)}_\sigma$
cannot fully contract to the $c$-ghost of the opposite string field for $n>2$.
Using the OPE  (61), (62), (75) it is straightforward to show
that for the string field $\Psi$ (33)-(35) the relevant terms 
in the BRST transformation are  given by:
\eqn\grav{\eqalign{Q\Psi\equiv{Q}{c}G({H}(B))=
\partial{c}c(B\partial_B{G(H(B))}-G(H(B)))\cr
+{1\over2}\partial^2{c}c
B\partial^2_B{G(H(B))}\cr
=\partial{c}(B\partial_B\Psi-\Psi)+{1\over2}\partial^2{c}
{B\partial_B^2\Psi}}}
with the notations explained above.
With all the above identities it is now straightforward to calculate
the SFT correlators.
The three-point correlator is then  computed to give
\eqn\grav{\eqalign{
<<g_1\circ{G}(H(B))(0)
g_2\circ{G}(H(B))(0)
g_3\circ{G}(H(B))(0)>>\cr
{\equiv}<<\Psi|\Psi\star\Psi>>
=T_1+T_2+T_3+T_4}}
Here
\eqn\grav{\eqalign{T_1=
\sum_{n_1,n_2,n_3=0}^\infty\sum_{m_1,m_2,m_3=0}^\infty
\sum_{q_1,q_2,q_3=0}^\infty
\sum_{k_1=0}^{m_1}\sum_{k_2=0}^{m_2}\sum_{k_3=0}^{m_3}
\sum_{N,R,T=0}^\infty
\cr
\sum_{N|n_1...n_{q_1}}\sum_{R|r_1...r_{q_2}}
\sum_{T|t_1...t_{q_3}}
\sum_{N_1=0}^{N+m_1-k_1}\sum_{R_1=0}^{R+m_2-k_2}\sum_{T_1=0}^{T+m_3-k_3}
\cr\lbrace
{{g_{q_1+1}g_{q_2+1}g_{q_3+1}}\over{q_1!q_2!q_3!}}h_{m_1}h_{m_2}h_{m_3}
h_{n_1}....h_{n_{q_1}}h_{r_1}...h_{r_{q_2}}
h_{t_1}....h_{t_{q_3}}
\cr\times
\prod_{i_1=1}^{q_2+1}\prod_{j_1=1}^{q_1+1}\prod_{i_2=1}^{q_3+1}\prod_{j_2=1}^{q_1+1}
\prod_{i_3=1}^{q_3+1}\prod_{j_3=1}^{q_2+1}
\lambda_{\alpha_{i_1j_1}|\beta_{j_1i_1}}
\lambda_{{\tilde{\alpha}}_{i_2j_2}|\beta_{j_2i_2}}
\lambda_{{\tilde{\alpha}}_{i_2j_2}|{\tilde{\beta}}_{j_2i_2}}
\cr
\prod_{\mu=1}^{q_1+1}
\sigma^{-1}(\lambda_{\alpha_{\mu,1}|\beta_{1,\mu}},...
\lambda_{\alpha_{\mu,{q_2+1}}|\beta_{{q_2+1},\mu}})
\sigma^{-1}(\lambda_{{\tilde{\alpha}}_{\mu,1}|\beta_{1,\mu}},...
\lambda_{{\tilde{\alpha}}_{\mu,{q_3+1}}|\beta_{{q_3+1},\mu}})
\cr
\prod_{\nu=1}^{q_2+1}
\sigma^{-1}(\lambda_{{\tilde{\alpha}}_{\mu,1}|{\tilde{\beta}}_{1,\mu}},...
\lambda_{{\tilde{\alpha}}_{\nu,{q_3+1}}|{\tilde{\beta}}_{{q_3+1},\nu}})
\cr\times
\sigma^{-1}(s_1^{(1)},...s_{q_2+1}^{(1)})
\sigma^{-1}({\tilde{s}}_1^{(1)},...{\tilde{s}}_{q_3+1}^{(1)})
\sigma^{-1}(s_1^{(2)},...s_{q_1+1}^{(2)})
\sigma^{-1}({\tilde{s}}_1^{(2)},...{\tilde{s}}_{q_3+1}^{(2)})
\sigma^{-1}(s_1^{(3)},...s_{q_1+1}^{(3)})
\cr
\sigma^{-1}({\tilde{s}}_1^{(3)},...{\tilde{s}}_{q_2+1}^{(3)})
((q_2+1)!(q_3+1)!)^{q_1+1}((q_3+1)!)^{q_2+1}
\cr\times
2^{N+T_1-N_1+m_1-k_1}({\sqrt{3}})^{N+R+T+m_1-k_1+m_2-k_2+m_3-k_3}
\rbrace
\cr
+permutations(g_1\circ\Psi,g_2\circ\Psi,g_3\circ\Psi)}}
where the exchange numbers are defined similarly to the previous case,
as well as the $\sigma^{-1}$-factors,
defined by products of array multiplicities in the relevant partitions.
Next, the $s,{\tilde{s}}$-numbers are similar to the $r,s$-numbers defined
previously and are related to conformal dimension losses
of string field components due to interactions
with partition elements (individual Bell polynomials) 
in  components of two opposite string fields.
 Altogether , these numbers satisfy the following constraints:
\eqn\grav{\eqalign{
\sum_{j=1}^{q_2+1}\alpha_{ij}+\sum_{j=1}^{q_3+1}{\tilde{\alpha}}_{ij}
=n_i(i=1,...,q_1+1)\cr
\sum_{j=1}^{q_1+1}\beta_{ij}+\sum_{j=1}^{q_3+1}{\tilde{\beta}}_{ij}
=r_i(i=1,...,q_2+1)\cr
\sum_{j=1}^{q_1+1}\gamma_{ij}+\sum_{j=1}^{q_2+1}{\tilde{\gamma}}_{ij}
=t_i(i=1,...,q_3+1)}}
and
\eqn\grav{\eqalign{
\sum_{i=1}^{q_1+1}\alpha_{ij}=s_j^{(1)}(j=1,...,q_2+1)\cr
\sum_{i=1}^{q_1+1}{\tilde{\alpha}}_{ij}={\tilde{s}}_j^{(1)}(j=1,...,q_3+1)\cr
\sum_{i=1}^{q_2+1}\alpha_{ij}=s_j^{(2)}(j=1,...,q_1+1)\cr
\sum_{i=1}^{q_2+1}{\tilde{\beta}}_{ij}={\tilde{s}}_j^{(2)}(j=1,...,q_3+1)\cr
\sum_{i=1}^{q_3+1}\gamma_{ij}=s_j^{(3)}(j=1,...,q_1+1)\cr
\sum_{i=1}^{q_2+1}{\tilde{\gamma}}_{ij}={\tilde{s}}_j^{(3)}(j=1,...,q_2+1)
}}
and  furthermore
\eqn\grav{\eqalign{
N+m_1-k_1=\sum_{i=1}^{q_1+1}n_i=\sum_{j=1}^{q_2+1}s_j^{(1)}
+\sum_{j=1}^{q_3+1}{\tilde{s}}_j^{(1)}\cr
R+m_2-k_2=\sum_{i=1}^{q_2+1}r_i=\sum_{j=1}^{q_1+1}s_j^{(2)}
+\sum_{j=1}^{q_3+1}{\tilde{s}}_j^{(2)}\cr
T+m_3-k_3=\sum_{i=1}^{q_3+1}t_i=\sum_{j=1}^{q_1+1}s_j^{(3)}
+\sum_{j=1}^{q_2+1}{\tilde{s}}_j^{(3)}}}
In other words, the exchange numbers, that form the OPE structure
of the Bell polynomial products, can be visualized as  
``partitions of partitions'' of the
conformal dimensions of the string field components.

This constitutes $T_1$, the first out of 4 terms contributing to the 
3-point correlator.
The remaining three can be obtained from $T_1$ by few simple 
replacements/manipulations.
That is, $T_2$ is obtained from $T_1$ by replacing
one of three $\alpha$ coefficients in (96) by the $\beta$-coefficient:
$\alpha(k_1,n_1)\rightarrow\beta(k_1,l_1|n_1)$,
with $k_1,l_1$ being summed over from 0 to $k_1+l_1=n_1$;
inserting an extra h-coefficient in the sum according to:
$h_{m_1}h_{m_2}h_{m_3}\rightarrow{h_{m_1}h_{m_2}h_{m_3}h_{m_4}}$,
replacing the difference $m_1-k_1\rightarrow{m_1+m_4-k_1-l_1}$
and finally replacing $q_1+1\rightarrow{q_1+2}$ in the upper limits
in the products over $j_1$ and $j_2$  in (96)-(99), as well as in the relevant
$g$-coefficient (the first among three in (96))
and in the relevant $\sigma^{-1}$-factors,
increasing their number of arguments by one unit - and finally,
permuting over the conformal transformations
 by $g_1$, $g_2$, $g_3$, as in $T_1$.
Thus the $T_1$-contribution has the $\alpha\alpha\alpha$-structure,
while $T_2$ carries the $\beta\alpha\alpha$-structure.
Similarly, to obtain $T_3$ out of $T_2$, one further
replaces the second $\alpha$-coefficient by the $\beta$-coefficient:
$\alpha(k_2,n_2)\rightarrow\beta(k_2,l_2|n_2)$, inserts an extra
$h$-coefficient in the product:
$h_{m_1}...h_{m_4}\rightarrow{h_{m_1}...h_{m_4}h_{m_5}}$,
and further replaces $m_2-k_2\rightarrow{m_2+m_5-k_2-l_2}$
and $q_2+1\rightarrow{q_2+2}$ according to the prescriptions explained above.
This, upon the permutation over the conformal transformations,
similar to the above, gives the $T_3$-contribution with the 
$\beta\beta\alpha$-structure.
The final contribution, $T_4$, having the $\beta\beta\beta$-structure,
is obtained similarly
from $T_3$ by replacing the last remaining $\alpha$ with $\beta$ and 
performing
the manipulations identical to those described above.
The overall expression for the three-point correlator  
thus looks complex enough.
Nevertheless, it is straightforward to check that, just as in the elementary
warm-up example demonstrated previously, the complicated sum given by 
(96)-(99)
can be converted successfully into the generating 
composite function and identified
with its series expansion.
Namely, we obtain:

\eqn\grav{\eqalign{
<<\Psi|\Psi\star\Psi>>\equiv
<<g_1\circ{G}(H(B))(0)
g_2\circ{G}(H(B))(0)
g_3\circ{G}(H(B))(0)>>
\cr
=
\sum_{j=1}^4{K_j}(G(H(F_\lambda)))}}
where
\eqn\grav{\eqalign{
{K_1}(G(H(F_\lambda)))=
\sum_{n_1=0}^\infty\sum_{n_2=0}^\infty\sum_{n_3=0}^\infty
\sum_{k_1=0}^{n_1}\sum_{k_2=0}^{n_2}\sum_{k_3=0}^{n_3}
\sum_{Q=1}^{k_1}\sum_{R=1}^{k_2}\sum_{S=1}^{k_3}
\cr\lbrace
\lbrack\alpha_{g_1}(k_1,n_1)\alpha_{g_2}(k_2,n_2)
\alpha_{g_3}(k_3,n_3)+
\alpha_{g_1}(k_1,n_1)\alpha_{g_3}(k_2,n_2)
\alpha_{g_2}(k_3,n_3)
\cr
+\alpha_{g_2}(k_1,n_1)\alpha_{g_1}(k_2,n_2)
\alpha_{g_3}(k_3,n_3)+
\alpha_{g_2}(k_1,n_1)\alpha_{g_3}(k_2,n_2)
\alpha_{g_1}(k_3,n_3)
\cr
+\alpha_{g_3}(k_1,n_1)\alpha_{g_2}(k_2,n_2)
\alpha_{g_1}(k_3,n_3)+
\alpha_{g_3}(k_1,n_1)\alpha_{g_1}(k_2,n_2)
\alpha_{g_2}(k_3,n_3)
\rbrace
\cr\times
\partial_{F_\lambda}^{k_1}H(F_\lambda(x,y))|_{x,y={\sqrt3}}
\partial_{F_\lambda}^{k_2}H(F_\lambda(x,y))|_{x,y={2\sqrt3}}
\partial_{F_\lambda}^{k_3}H(F_\lambda(x,y))|_{x,y={\sqrt3}}
\cr\times
G^\prime(\partial^QH(F_\lambda(x,y)))|_{x,y={\sqrt{3}}}
G^\prime(\partial^RH(F_\lambda(x,y)))|_{x,y={2\sqrt{3}}}
G^\prime(\partial^SH(F_\lambda(x,y)))|_{x,y={\sqrt{3}}}
}}
\eqn\grav{\eqalign{
{K_2}(G(H(F_\lambda)))=
\sum_{n_1=0}^\infty\sum_{n_2=0}^\infty\sum_{n_3=0}^\infty
\sum_{k_1,l_1=0}^{k_1+l_1=n_1}\sum_{k_2=0}^{n_2}\sum_{k_3=0}^{n_3}
\sum_{Q=1}^{k_1+l_1-1}\sum_{R=1}^{k_2}\sum_{S=1}^{k_3}
\cr
\beta_{g_1}(k_1,l_1|n_1)
(\alpha_{g_2}(k_2,n_2)
\alpha_{g_3}(k_3,n_3)+\alpha_{g_3}(k_2,n_2)
\alpha_{g_2}(k_3,n_3))
\cr\times
G^{\prime\prime}(\partial^QH(F_\lambda(x,y)))|_{x,y={\sqrt{3}}}
G^\prime(\partial^RH(F_\lambda(x,y)))|_{x,y={2\sqrt{3}}}
G^\prime(\partial^SH(F_\lambda(x,y)))|_{x,y={\sqrt{3}}}
\cr\times
(\partial_{F_\lambda}^{k_1}H(F_\lambda(x,y))|_{x,y={\sqrt3}}
\partial_{F_\lambda}^{l_1}H(F_\lambda(x,y))|_{x,y={\sqrt3}}
\cr\times
\partial_{F_\lambda}^{k_2}H(F_\lambda(x,y))|_{x,y={2\sqrt3}}
\partial_{F_\lambda}^{k_3}H(F_\lambda(x,y))|_{x,y={\sqrt3}})
\cr
+
\beta_{g_2}(k_1,l_1|n_1)
(\alpha_{g_1}(k_2,n_2)
\alpha_{g_3}(k_3,n_3)+\alpha_{g_3}(k_2,n_2)
\alpha_{g_1}(k_3,n_3))
\cr\times
G^{\prime\prime}(\partial^QH(F_\lambda(x,y)))|_{x,y={2\sqrt{3}}}
G^\prime(\partial^RH(F_\lambda(x,y)))|_{x,y={\sqrt{3}}}
G^\prime(\partial^SH(F_\lambda(x,y)))|_{x,y={\sqrt{3}}}
\cr\times
(\partial_{F_\lambda}^{k_1}H(F_\lambda(x,y))|_{x,y={2\sqrt3}}
\partial_{F_\lambda}^{l_1}H(F_\lambda(x,y))|_{x,y={2\sqrt3}}
\cr\times
\partial_{F_\lambda}^{k_2}H(F_\lambda(x,y))|_{x,y={\sqrt3}}
\partial_{F_\lambda}^{k_3}H(F_\lambda(x,y))|_{x,y={\sqrt3}})
\cr
+
\beta_{g_3}(k_1,l_1|n_1)
(\alpha_{g_1}(k_2,n_2)
\alpha_{g_2}(k_3,n_3)+\alpha_{g_2}(k_2,n_2)
\alpha_{g_1}(k_3,n_3))
\cr\times
G^{\prime}(\partial^QH(F_\lambda(x,y)))|_{x,y={\sqrt{3}}}
G^\prime(\partial^RH(F_\lambda(x,y)))|_{x,y={\sqrt{3}}}
G^\prime(\partial^SH(F_\lambda(x,y)))|_{x,y={2\sqrt{3}}}
\cr\times
(\partial_{F_\lambda}^{k_1}H(F_\lambda(x,y))|_{x,y={\sqrt3}}
\partial_{F_\lambda}^{l_1}H(F_\lambda(x,y))|_{x,y={\sqrt3}}
\cr\times
\partial_{F_\lambda}^{k_2}H(F_\lambda(x,y))|_{x,y={\sqrt3}}
\partial_{F_\lambda}^{k_3}H(F_\lambda(x,y))|_{x,y={2\sqrt3}})
}}
\eqn\grav{\eqalign{{K_3}(G(H(F_\lambda)))=
\sum_{n_1=0}^\infty\sum_{n_2=0}^\infty\sum_{n_3=0}^\infty
\sum_{k_1,l_1=0;k_1<l_1}^{k_1+l_1=n_1}\sum_{k_2,l_2=0;k_2<l_2}^{k_2+l_2=n_2}\sum_{k_3=0}^{n_3}
\sum_{Q=1}^{k_1+l_1-1}\sum_{R=1}^{k_2+l_2-1}\sum_{S=1}^{k_3}
\cr
(\beta_{g_1}(k_1,l_1|n_1)\beta_{g_2}(k_2,,l_2|n_2)+
\beta_{g_2}(k_1,l_1|n_1)\beta_{g_1}(k_2,,l_2|n_2))
\alpha_{g_3}(k_3,n_3)
\cr\times
G^{\prime\prime}(\partial^QH(F_\lambda(x,y)))|_{x,y={\sqrt{3}}}
G^{\prime\prime}(\partial^RH(F_\lambda(x,y)))|_{x,y={2\sqrt{3}}}
G^\prime(\partial^SH(F_\lambda(x,y)))|_{x,y={\sqrt{3}}}
\cr\times
(\partial_{F_\lambda}^{k_1}H(F_\lambda(x,y))|_{x,y={\sqrt3}}
\partial_{F_\lambda}^{l_1}H(F_\lambda(x,y))|_{x,y={\sqrt3}}
\partial_{F_\lambda}^{k_2}H(F_\lambda(x,y))|_{x,y={2\sqrt3}}
\cr\times
\partial_{F_\lambda}^{l_2}H(F_\lambda(x,y))|_{x,y={2\sqrt3}})
\partial_{F_\lambda}^{k_3}H(F_\lambda(x,y))|_{x,y={\sqrt3}}
\cr
+
(\beta_{g_1}(k_1,l_1|n_1)\beta_{g_3}(k_2,,l_2|n_2)+
\beta_{g_3}(k_1,l_1|n_1)\beta_{g_1}(k_2,,l_2|n_2))
\alpha_{g_2}(k_3,n_3)
\cr\times
G^{\prime\prime}(\partial^QH(F_\lambda(x,y)))|_{x,y={2\sqrt{3}}}
G^{\prime\prime}(\partial^RH(F_\lambda(x,y)))|_{x,y={\sqrt{3}}}
G^\prime(\partial^SH(F_\lambda(x,y)))|_{x,y={\sqrt{3}}}
\cr\times
(\partial_{F_\lambda}^{k_1}H(F_\lambda(x,y))|_{x,y={2\sqrt3}}
\partial_{F_\lambda}^{l_1}H(F_\lambda(x,y))|_{x,y={2\sqrt3}}
\partial_{F_\lambda}^{k_2}H(F_\lambda(x,y))|_{x,y={\sqrt3}}
\cr\times
\partial_{F_\lambda}^{l_2}H(F_\lambda(x,y))|_{x,y={\sqrt3}})
\partial_{F_\lambda}^{k_3}H(F_\lambda(x,y))|_{x,y={\sqrt3}}
\cr
+
(\beta_{g_2}(k_1,l_1|n_1)\beta_{g_3}(k_2,,l_2|n_2)+
\beta_{g_3}(k_1,l_1|n_1)\beta_{g_2}(k_2,,l_2|n_2))
\alpha_{g_1}(k_3,n_3)
\cr\times
G^{\prime\prime}(\partial^QH(F_\lambda(x,y)))|_{x,y={\sqrt{3}}}
G^{\prime\prime}(\partial^RH(F_\lambda(x,y)))|_{x,y={\sqrt{3}}}
G^\prime(\partial^SH(F_\lambda(x,y)))|_{x,y={2\sqrt{3}}}
\cr\times
(\partial_{F_\lambda}^{k_1}H(F_\lambda(x,y))|_{x,y={\sqrt3}}
\partial_{F_\lambda}^{l_1}H(F_\lambda(x,y))|_{x,y={\sqrt3}}
\partial_{F_\lambda}^{k_2}H(F_\lambda(x,y))|_{x,y={\sqrt3}}
\cr\times
\partial_{F_\lambda}^{l_2}H(F_\lambda(x,y))|_{x,y={\sqrt3}})
\partial_{F_\lambda}^{k_3}H(F_\lambda(x,y))|_{x,y={2\sqrt3}}
}}

\eqn\grav{\eqalign{{K_4}(G(H(F_\lambda)))
\cr
=
\sum_{n_1=0}^\infty\sum_{n_2=0}^\infty\sum_{n_3=0}^\infty
\sum_{k_1,l_1=0;k_1<l_1}^{k_1+l_1=n_1}\sum_{k_2,l_2=0;k_2<l_2}^{k_2+l_2=n_2}
\sum_{k_3,l_3=0;k_3<l_3}^{n_3}
\sum_{Q=1}^{k_1+l_1-1}\sum_{R=1}^{k_2+l_2-1}\sum_{S=1}^{k_3+l_3-1}
\cr
(\beta_{g_1}(k_1,l_1|n_1)\beta_{g_2}(k_2,l_2|n_2)\beta_{g_3}(k_3,l_3|n_3)
+
\beta_{g_1}(k_1,l_1|n_1)\beta_{g_3}(k_2,l_2|n_2)\beta_{g_2}(k_3,l_3|n_3)
\cr
+
\beta_{g_1}(k_2,l_2|n_2)\beta_{g_2}(k_1,l_1|n_1)\beta_{g_3}(k_3,l_3|n_3)
+
\beta_{g_1}(k_2,l_2|n_2)\beta_{g_3}(k_1,l_1|n_1)\beta_{g_2}(k_3,l_3|n_3)
\cr
+
\beta_{g_1}(k_3,l_3|n_3)\beta_{g_2}(k_1,l_1|n_1)\beta_{g_3}(k_2,l_2|n_2)
+
\beta_{g_1}(k_3,l_3|n_3)\beta_{g_3}(k_1,l_1|n_1)\beta_{g_2}(k_2,l_2|n_2)
)
\cr\times
G^{\prime\prime}(\partial^QH(F_\lambda(x,y)))|_{x,y={\sqrt{3}}}
G^{\prime\prime}(\partial^RH(F_\lambda(x,y)))|_{x,y={\sqrt{3}}}
G^{\prime\prime}(\partial^SH(F_\lambda(x,y)))|_{x,y={2\sqrt{3}}}
\cr\times
(\partial_{F_\lambda}^{k_1}H(F_\lambda(x,y))|_{x,y={\sqrt3}}
\partial_{F_\lambda}^{l_1}H(F_\lambda(x,y))|_{x,y={\sqrt3}}
\partial_{F_\lambda}^{k_2}H(F_\lambda(x,y))|_{x,y={\sqrt3}}
\cr\times
\partial_{F_\lambda}^{l_2}H(F_\lambda(x,y))|_{x,y={\sqrt3}}
\partial_{F_\lambda}^{k_3}H(F_\lambda(x,y))|_{x,y={2\sqrt3}}
\partial_{F_\lambda}^{k_3}H(F_\lambda(x,y))|_{x,y={2\sqrt3}}
)
}}
where, in our notations, $G(\partial^Q{H})$ is obtained
from $G(H)$ by replacing the argument $H\rightarrow\partial^{Q}H$
and $\partial^Q{H}\equiv{{\partial^Q{H}}\over{\partial{{F_\lambda}^Q}}}$
 and finally
$G^{\prime}(\partial^{Q}H)={{\partial{G}}\over{\partial(\partial^Q{H})}}$.
This concludes the computation of the three-point SFT correlator 
for our solution ansatz.
The final step is to compute the kinetic term
$<<Q\Psi(0)|I\circ\Psi(0)>>$ using the operator products  
and the identities derived above.
According to the BRST transformation identity (94)
this correlator is  determined by two contributions:
one proportional to the ghost part 
$<c(z)\partial{c}c(w>|_{z=0;w\rightarrow\infty}=(z-w)^2$
another to 
$<c(z)\partial^2{c}c(w)>|_{z=0;w\rightarrow\infty}=-2(z-w)$.
Note that, since $c$, $\partial{c}c$  and $\partial^2c{c}$ ghost  fields
have conformal dimensions $-1$, $-1$ and $0$ respectively,
and since the conformal transformation by $I(z)$ takes $0$ to infinity,
it is straightforward
to check that the matter part of the first contribution
only contains
the terms with the conformal dimensions of the string field components
at $z$ equal to those of the string components at $w$;
all the terms with unequal conformal dimensions of operators
at $z$ and $w$ vanish
in the limit $w\rightarrow\infty$. Similarly, 
the matter part of  the second contribution (multiplied
by the $<c(z)\partial^2{c}c(w)>|_{z=0;w\rightarrow\infty}$ ghost correlator)
only contains the terms with the conformal dimensions of the operators
at $z$ equal to those of the operators at $w$ plus one.

Then, performing straightforward  calculation  
of the correlator, similar to those above,
 plugging into SFT equations of motion (1)
 leads to the  defining relation for the
$G(H(F_\lambda))$ function of our ansatz, given by:
\eqn\grav{\eqalign{
\sum_{n=0}^\infty\sum_{k=1}^n
\alpha_I(k,n)\sum_{Q=0}^k
{\lbrace}Z_0\lbrack
{\tilde{G}}(H(F_\lambda))|G^\prime(\partial^Q_{F_\lambda}H(F_\lambda))\rbrack
\cr
-Z_1\lbrack
\partial_{F_\lambda}{\tilde{G}}(H(F_\lambda))|G^\prime(\partial^Q_{F_\lambda}
{\tilde{H}}(F_\lambda))\rbrack\rbrace
\rbrack\partial^k_{F_\lambda}{\tilde{H}}(F_\lambda)|_{x=y=1}
\cr
+\sum_{n=0}^\infty
\sum_{k,l=1;k<{l}}^{n}\beta_I(k,l|n)\sum_{Q=0}^{k+l-1}
\rbrace
Z_0\lbrack{\tilde{G}}({\tilde{H}}(F_\lambda))|G^{\prime\prime}(\partial^Q_{F_\lambda}
{\tilde{H}}(F_\lambda))\rbrack
\cr
-
Z_1\lbrack
\partial_{F_\lambda}{\tilde{G}}(H(F_\lambda))|G^{\prime\prime}(\partial^Q_{F_\lambda}
{\tilde{H}}(F_\lambda))\rbrack\rbrace
\partial^k_{F_\lambda}{\tilde{H}}(F_\lambda)|\partial^l_{F_\lambda}
{\tilde{H}}(F_\lambda)|_{x,y=1}
\rbrace
\cr
=\sum_j{K_j(G(H(F_\lambda)))}
}}
where the operations $Z_k{\lbrack}f_1(x)|f_2(x)\rbrack$ 
acting on functions $f_1$ and $f_2$ 
are defined as follows:
if $f_1(x)=\sum_{m}a_m{x^m}$ and $f_2(x)=\sum_{n}a_nx^n$
are the series expansions for $f_1$ and $f_2$
then $Z_k$ maps them into the function (formal series)
\eqn\grav{\eqalign{
Z_k{\lbrack}f_1(x)|f_2(y)\rbrack=\sum_{n}a_n{b_{n+k}}x^ny^{n+k}
}}
The tilde operations are again defined according to:
\eqn\grav{\eqalign{
{\tilde{G}}(H(x))=x{{d}\over{dx}}G(H(x))-G(H(x))\cr
{\tilde{H}}(x)=x{{d}\over{dx}}H(x)-H(x)}}
and the $K_j$-functions are defined by (101)-(104).
The functional equation (105) is the main result of this paper
and  constitutes the defining relation for the SFT ansatz.
As cumbersome as this relation is, it can be, e.g., solved order by order 
by iterations and reduces the SFT equation (1) to the identity
which is essentially algebraic.
In the case of $\alpha^2=-2$ that we mostly have explored in this paper,
one particularly simple example for the
generating functions solving the defining relation (105)
is given by
\eqn\grav{\eqalign{
G(H)={1\over{1-H(F_\lambda)}}\cr
H(x)={1\over{1-x}}}}
Replacing $x^{n}\rightarrow{B_{\vec{\psi}}^{(n)}}$ according to our usual
prescription, leads to the generating function of
the SU(2) enveloping algebra $T(\mu)$ with the parameter $\mu$
 defined by the $F_\lambda$ (elementary correlator of Bell polynomial
operators). The classical $w_\infty$ algebra is then recovered in the
simple pole $\sim{(z-w)^{-1}}$ of the OPE of $G(H(B^{(n)}(z)))$
and $G(H(B^{(n)}(w)))$.
In general, the defining relation (105) appears to parametrize the class
of SFT solutions, of which (108) is an elementary example.
Finding the explicit form of this class of the solutions ,
generalizing (108) appears to be
an important challenge and obviously doesn't seem to be easy.
However, it appears that this class
is  most naturally  expressible in terms 
of the series in the powers of the generating function (108) for
the products of the Bell polynomial operators,
relating it to the enveloping of the enveloping of $SU(2)$
(and more particularly, to the enveloping of $w_\infty$)
The objects like these are known to be relevant to the quantization
of higher spin theories and to the multi-particle realizations of the
higher-spin algebras  ~{\vmulti}.
The crucial point about the SFT solutions,
constrained to the subspace of operators given by products of the Bell
polynomials, is that these objects

a) behave in a controllable and consistent way in the SFT star product
computations

b)form a natural operator basis for the free-field
realization of the $SU(2)$ envelopings and $w_\infty$.
 In the concluding section,
we shall briefly discuss how the construction, studied in this paper,
can be generalized to higher space-time dimensions.

\centerline{\bf Conclusion and Discussion}

In this  paper we have considered the ansatz solution in bosonic
string field theory, given by formal series in partial (incomplete)
Bell polynomials of Bell polynomial operators in the worldsheet
derivatives of the target space fields. These objects form an operator
algebra realization for the enveloping of $SU(2)$, including
the $w_\infty$ algebra appearing at the simple pole  of the OPE.
This, up to
ideal factorization, is isomorphic to a chiral copy of higher spin algebra in
$AdS_3$. 
The solution is given in terms of
of the functional constraints on the generating 
functions for the operators realizing this enveloping.
These constraints altogether are  quite  cumbersome
and finding their manifest solutions doesn't appear to be an easy challenge,
except for a relatively simple example (108).

 Nevertheless, the constraints 
for the $h$ and $g$-expansion coefficients are essentially algebraic and
 in principle be  analyzed order by order by iterations.

An important question is whether the construction, considered in our work,
can be extended to SFT solutions involving higher-dimensional 
 enveloping/higher spin algebras.
A possible answer to that may come from superstring generalization of
the computation performed in this work and switching on the $\beta-\gamma$
system  of the superconformal ghosts.
Just as the solution, considered in this paper, was in a sense inspired 
by the bosonic $c=1$ model (an elementary example pointing out the relevance 
of operator algebra involving Bell polynomial products to $w_\infty$ and 
 higher spin algebra)
one can use a supersymmetric $c=1$ model coupled to the $\beta-\gamma$ 
system as a toy model
inspiration. It is known  that the interaction with the superconformal ghosts
enhances the $SU(2)$ symmetry at the selfdual point to $SU(N)$
where $N-2$ is the maximal superconformal ghost number (ghost cohomology rank)
of the generators ~{\selflast} 
One can hope that manifest form of the vertex operators in this model
would prompt us the form of the ansatz we should be 
looking for, and the resulting
solution would be relevant to envelopings of $SU(N)$ 
or their subalgebras, related
to isometries of $AdS$ in different space-time dimensions.
The manifest form of the vertex operators in this model would
again involve the products of Bell polynomials,
 however their structure will be 
far more diverse. In the present paper the 
$\psi={\vec{\alpha}}{\vec{X}}$ parameter
of the operators was fixed to be the same for 
the all the string field components
(as it is the same for all the operators for the bosonic discrete states 
and is equal to
$-i{\sqrt{2}}X$). Switching on the higher superconformal ghost pictures 
in the $c=1$
model would then result
in the appearance of Bell polynomial products with mixed $\psi$-parameters.
While the naive number of the parameters would be ${1\over2}N(N-1)$ 
(total number
of the lowering operators of $SU(N)$), the actual number
 would be less and of the order of $N$,
since not all the lowering operators, acting on tachyonic primaries, 
lead to physically
distinct states. The distinct states
 are basically generated by the
lowering operators of ghost numbers $N-2$ carrying the maximum momentum value
in the $X$-direction, equal to $N-1$ , and the total number of such generators
is $N-1$. Thus one can hope that introducing extra $\psi$-parameters
will direct us towards the SFT solutions describing the higher-dimensional
enveloping/ higher spin algebras. It looks plausible that the framework
 involving the WZW-type  Berkovits
string field theory may turn out to be  a convenient framework for this 
program along with
cubic superstring field theory with picture-changing insertions 
~{\witsfts, \iaf, \ias, \yost}.
Following this strategy, one can hope to find the defining 
constraints for the generating functions,
similar to those considered in this paper.
It would certainly be of interest and of
importance to study these constraints and to identify
some of their manifest solutions.
This hopefully shall lead to new important insights 
regarding nonperturbative higher spin configurations, 
as well as to deeper understanding 
of the underlying relations between SFT and higher 
spin field theories, which appear to be crucial
ingredients of holography principle in general.

\centerline{\bf Acknowledgements}

It is a great pleasure to thank the speakers and the
 participants of the annual string field
theory conference (SFT-2015) in Chengdu 
(May 11-16, 2015) , as well as the speakers at the introductory school,
for the illuminating lectures, talks and productive discussions.
It is also a great pleasure to thank Bo Ning, Zheng Sun and Haitang Yang
for our collaboration in organizing SFT-2015.

\listrefs

\end